\documentclass[10pt,journal,twocolumn]{IEEEtran}
\makeatletter
\def\blfootnote{\xdef\@thefnmark{}\@footnotetext}
\def\ps@headings{%
\def\@oddhead{\mbox{}\scriptsize\rightmark \hfil \thepage}%
\def\@evenhead{\scriptsize\thepage \hfil \leftmark\mbox{}}%
\def\@oddfoot{}%
\def\@evenfoot{}}

\makeatother

\usepackage{ifthen}
\def\Context{}
\def\Context{FULLPAPER}

\ifthenelse{\equal{\Context}{FULLPAPER}}{
    \pagestyle{headings}
    \newcommand{\TR}[1]{{#1}}
    \newcommand{\TON}[1]{}
}
{
    \pagestyle{empty}
    \newcommand{\TR}[1]{}
    \newcommand{\TON}[1]{{#1}}
}

\usepackage[top=0.75in, bottom=1in, left=0.625in, right=0.625in]{geometry}

\ifCLASSINFOpdf
  \usepackage[pdftex]{graphicx}
  \usepackage{epstopdf}

   \graphicspath{{./figure/}{./}}
  \DeclareGraphicsExtensions{.pdf,.jpeg,.png}
\else
  \usepackage[dvips]{graphicx}
     \graphicspath{{./figure/}}
\fi
\usepackage{caption}
\usepackage{subcaption}
\usepackage{bbm}
\usepackage[cmex10]{amsmath}
\usepackage{amssymb}
\usepackage{verbatim}
\usepackage{tipa}
\usepackage{stmaryrd}
\usepackage{fancybox}
\usepackage[hidelinks]{hyperref}
\usepackage[usenames, dvipsnames]{color}

\usepackage{url}

\usepackage{leading}
\leading{12pt}

\usepackage{color}
\usepackage[usenames,dvipsnames,svgnames,table]{xcolor}
\definecolor{light-gray}{gray}{0.9}

\usepackage[ruled,,figure]{algorithm2e}
\SetKwInput{KwVar}{Variables}
\SetKw{KwStage}{Stage}
\SetAlgoNlRelativeSize{0}
\SetNlSty{textbf}{}{)}
\SetNlSkip{0em}
\newtheorem{theorem}{Theorem}[section]
\newtheorem{lemma}{Lemma}[section]
\newtheorem{corollary}{Corollary}[section]

\newtheorem{definition}{Definition}[section]

\newcommand{\ignore}[1]{}

\newcommand{\Step}{\emph}
\newcommand{\comments}[1]{}

\newlength{\DepthReference}
\newlength{\HeightReference}
\newlength{\Width}%

\newcommand{\MyColorBox}[2][light-gray]%
{%
    \settowidth{\Width}{#2}%
    \colorbox{#1}%
    {%
        \raisebox{-\DepthReference}%
        {%
                \parbox[b][\HeightReference+\DepthReference][c]{\Width}{\centering#2}%
        }%
    }%
}

\title{Tunable QoS-Aware Network Survivability}

\TR{
\author{\IEEEauthorblockN{Jose Yallouz}
\IEEEauthorblockA{Department of Electrical Engineering\\Technion, Israel Institute of Technology\\
Email: jose@tx.technion.ac.il}
\and
\IEEEauthorblockN{Ariel Orda}
\IEEEauthorblockA{Department of Electrical Engineering\\Technion, Israel Institute of Technology\\
Email: ariel@ee.technion.ac.il}}
}
\TON{\author{\IEEEauthorblockN{\large Jose Yallouz and Ariel Orda}}}

\begin{document}             
\maketitle                   

\begin{abstract}
Coping with network failures has been recognized as an issue of major importance in terms of social security, stability and prosperity. It has become clear that current networking standards fall short of coping with the complex challenge of surviving failures. The need to address this challenge has become a focal point of networking research. In particular, the concept of \textbf{\emph{tunable survivability}} offers major performance improvements over traditional approaches. Indeed, while the traditional approach aims at providing full (100\%) protection against network failures through disjoint paths, it was realized that this requirement is too restrictive in practice. Tunable survivability provides a quantitative measure for specifying the desired level (0\%-100\%) of survivability and offers flexibility in the choice of the routing paths. Previous work focused on the simpler class of ``bottleneck'' criteria, such as bandwidth. In this study, we focus on the important and much more complex class of \emph{additive} criteria, such as delay and cost. First, we establish some (in part, counter-intuitive) properties of the optimal solution. Then, we establish efficient algorithmic schemes for optimizing the level of survivability under additive end-to-end QoS bounds. Subsequently, through extensive simulations, we show that, at the price of \emph{negligible} reduction in the level of survivability, a major improvement  (up to a factor of $2$) is obtained in terms of end-to-end QoS performance. Finally, we exploit the above findings in the context of a network design problem, in which, for a given investment budget, we aim to improve the survivability of the network links.
\end{abstract}


\section{Introduction}
\label{sec:intro}
\TON{
\blfootnote{
J. Yallouz and A. Orda are with the Department of Electrical Engineering, Technion, Haifa 320000,
Israel (e-mails: \{jose@tx, ariel@ee\}.technion.ac.il).
}\par
}

The internet infrastructure has been progressing rapidly since its deployment.
{Nowadays technologies offer rates of 100 Gbit/s and beyond \cite{IEEEP802.3ba, Christopher:R2011472}.} Current core routers, such as CRS-3, reach capacities of hundreds of terabits per second {\cite{CRS-3}}. With this extreme increase of transmission rates, any failure in the network infrastructure{, e.g. a fiber cut or a router shutdown,}  may lead to a vast amount of data loss. Hence, survivability in the network is becoming increasingly important.

In particular, failures in the network infrastructure should be recovered promptly.
For example, some standard recommendations, e.g. \cite{G.8032} \cite{rfc6372}, require that recovery from a single failure should be performed within $50 ms$. The literature distinguishes between two major classes of recovery schemes, namely restoration and protection \cite{askarian2008protection}. In \emph{restoration schemes}, post-failure actions are performed in order to search for a backup path that would avoid the faulty element. In \emph{protection schemes}, on the other hand, pre-failure actions are performed in order to pre-establish a backup solution for any possible failure.
Protection schemes have an obvious advantage in terms of recovery time and are usually achieved by the establishment of pairs of disjoint paths.
{
Specifically, protection schemes have been implemented in several network architectures, e.g. SONET/SDH and MPLS. In  \emph{Multi-Protocol Label Switching} (MPLS) \cite{rfc3031}, two major protection schemes are employed, namely $\textsc{1:1}$ and $\textsc{1+1}$. In $\textsc{1:1}$ protection, the data is sent only over a single path, while the backup path is activated upon a failure on the first path. In $1+1$ protection, the data is duplicated over both paths.
}

{We adopt the widely used single link failure model that aims at handling  single failure events.  This model has been the focus of numerous studies on survivability, e.g. \TR{\cite{alicherry2004preprovisioning} \cite{Frederick2003} \cite{ramamurthy2003survivable}} \cite{BejeranoBORS05} \cite{banner2010designing} \cite{tapolcai2009monitoring}.
While the case of multiple failures should be considered, and, indeed, has been the subject of several studies (e.g., \cite{bhattacharya2011qos, johnston2011robust, heegaard2009network}),\footnote{{Some of these studies, e.g. \cite{heegaard2009network}, considered the failure of multiple components due to a single fault.}}  the single failure model does merit attention, due to several reasons. First, when exploring novel survivability schemes, it is natural to begin with this basic case, whose analysis would then provide insight for future enhancements for handling multiple failures.
Moreover, protecting against a single failure is a common requirement of several survivability standards, e.g. \cite{G.8032} \cite{rfc6372}.
In addition, a common approach for handling multiple failures is to supply protection for the first failure and restoration for any subsequent ones.
Moreover, being a first step in proposing a novel scheme, in this study we focus on single independent failures, such as fiber cuts and router shutdown, while enhancements for the case of dependent failures (e.g. due to cyber attacks and natural disasters) remain an important subject for future work.
}

Under the single link failure model, the employment of disjoint paths provides full (100\%) protection. Hence, this is the common solution approach of path protection schemes.
However, the requirement of fully link-disjoint paths is often too restrictive and demands excessive redundancy
in practice. Furthermore, a pair of disjoint paths of sufficient quality may not exist,
occasionally making the requirement infeasible. Therefore, a milder and more flexible survivability concept is called for, which would relax the rigid requirement of link-disjoint paths by also considering paths containing common links. Accordingly, a previous study \cite{Banner:2007} introduced the novel concept of \emph{tunable survivability},
which provides a quantitative  measure to specify the desired level of survivability. This concept allows any
degree of survivability in the range 0\% to 100\%, thus transforming survivability into a quantifiable metric.

Specifically, tunable survivability enables the establishment of connections that can survive network
failures with any desired probability. Given a connection that consists of two paths \footnote{{As shall be explained in the sequel, we include the case of two identical paths.}}
between a source-destination pair under the single failure model, only a failure on a link that is common to both paths can disrupt the connection. Accordingly, we characterize a connection as $p$-survivable if there is a probability of at least $p$ to have all common links operational.

{\emph{Quality of Service (QoS)} refers to the capability of a network to provide guarantees to deliver predictable results \cite{QoS}.}
{Elements of network performance within the scope of QoS metrics often include survivability, bandwidth, delay, jitter and cost.} 
Generally, we distinguish between two classes of QoS metrics, namely: bottleneck metrics, such as bandwidth, which are defined by the weakest component in the path, and additive metrics, such as delay, which are defined by the sum of the corresponding metrics over the path's links.
Algorithmic schemes that combine the concept of tunable survivability with bottleneck metrics were established in \cite{Banner:2007} {and \cite{YallouzRO14}}.
However, the important and much more complex class of additive metrics was not considered. Accordingly, this is the subject of the present work.

{
When a connection is composed of two paths, there are several possibilities for defining its weight out of the weights of the connection's paths. Indeed, various studies considered several weight definitions  in the context of connections based on link-disjoint paths.
A natural choice is to consider  the minimum of the lengths of the two paths.
However, this approach results in strongly NP-complete problems \cite{xu2006complexity}, namely even approximate solutions
are computationally intractable.  Alternatively, we can consider the worst (highest) among the weights of the two paths,
yet this also leads to an NP-Hard problem \cite{li1990complexity}. 
Finally, a common approach  is to consider the sum of the lengths of the two paths, which attempts to minimize the aggregate weight of the two paths (e.g., \cite{Suurballe:1984}, \cite{4215794}).
Beyond allowing computationally efficient optimal solutions, we shall indicate that this approach also provides a $2$-approximation solution to the previous approach, which targets at minimizing the higher length of the two paths. 
}

\begin{figure}[tb]
  \centering
    \includegraphics[width=0.48\textwidth]{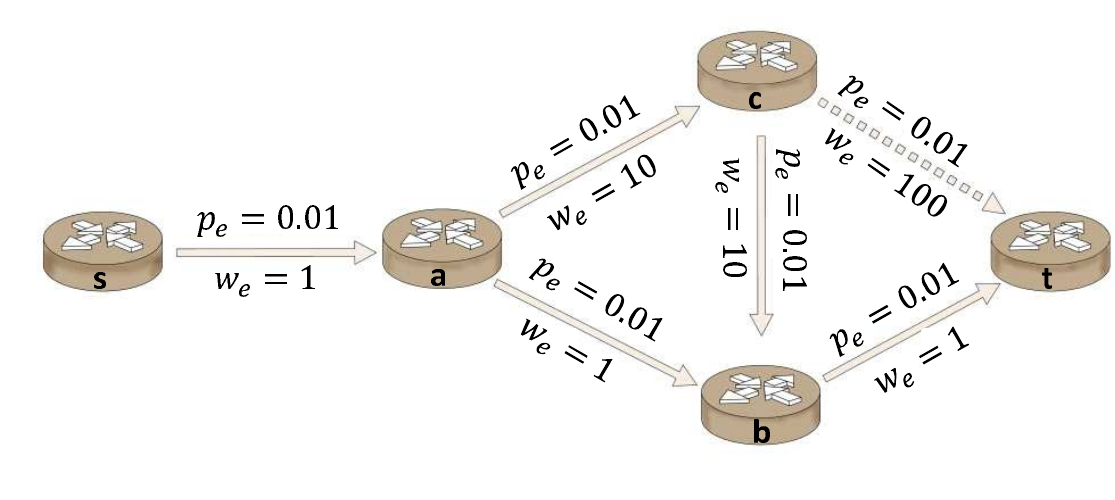}
  \caption{Example of $p$-survivable connections under an additive QoS metric}
\label{fig:intro_example}
\end{figure}

The following example demonstrates the concept of $p$-survivable connections combining an additive QoS metric and its advantages over traditional protection schemes. Consider the network described in Fig. \ref{fig:intro_example}, where each link is associated with a failure probability $p_e$ and a weight $w_e$ representing an additive metric. Assume that a connection is to be established between $s$ and $t$. Here, the weight of a $p$-survivable connection is
defined as the sum of all the weights of the connection's links, considering the weight of a link that is common to both paths only once.
As no pair of disjoint paths from $s$ to $t$ exists, there is no full protection against single failures, and the traditional
survivability requirement is thus infeasible. However, if we are satisfied with $0.99$-survivability against single network failures, then a connection that consists of the paths $\pi_{1}=<s,a,c,t>$ and $\pi_{2}=<s,a,b,t>$ is a valid solution, since the only (single) failure that can concurrently damage both paths is in the common link $(s,a)$. Hence, as this link fails with a probability of $0.01$, the connection is $0.99$-survivable. The weight of the connection, defined by the sum of all link weights $w_{e}$ (counting the weights of the common links
just once), is $\mathbf{113}$. Now, suppose that we are satisfied with $0.99^2$-survivability. Clearly, the paths $\pi_{1}=<s,a,c,b,t>$ and $\pi_{2}=<s,a,b,t>$ also constitute a valid connection, for which the weight decreases to $\mathbf{23}$. Finally, assume that we are satisfied with $0.99^3$-survivability. {Now, the (identical) paths $\pi_{1}=<s,a,b,t>$ and $\pi_{2}=<s,a,b,t>$  also become feasible,} thus decreasing the weight of the connection to $\mathbf{3}$.

Motivated by \cite{Banner:2007}, we investigate how to combine the tunable survivability concept with additive QoS guarantees. To that end, in Section \ref{sec:formulation}, we formulate an optimization problem that considers two requirements, namely a (minimum) level of survivability and an additive end-to-end QoS guarantee.
We establish some fundamental properties of the structure of the optimal solution. In particular, in Section \ref{sec:ct_charc}, we prove that, for an important class of problems, only a (typically small) subset of the network's links may affect the survivability value of the optimal solution. Next, in Section \ref{sec:Intractability}, we establish that our class of problems is computationally intractable. Accordingly, in Section \ref{sec:algo}, we design and validate a pseudo-polynomial solution and an efficient Fully Polynomial-Time Approximation Scheme (FPTAS).
In Section \ref{sec:simulation}, through comprehensive simulations, we show that, typically, a modest relaxation (of a few percents) in the survivability level is enough to provide a major improvement in terms of the QoS requirement, e.g. cutting by half the end-to-end delay.
Then, in Section \ref{sec:network_design}, we exploit the above findings in the context of a network design problem, in which we need to best invest a given ``budget'' for improving the survivability of the network links. {In Setion \ref{sec:min-max}, we show that the algorithmic scheme presented in Section \ref{sec:algo} provides a $2$-approximation for an intractable variant of the problem.}
Finally, Section \ref{sec:conclusions} summarizes our results and discusses directions for future research. \TON{Due to space limits some details are omitted from this version and can
be found (online) in \cite{TR-QoS-Aware-Survivability}.}

\section{Model and Problem Formulation}
\label{sec:formulation}
A \emph{network} is represented by a directed graph $G(V,E)$, where $V$ is the set of nodes and $E$ is the set
of links. We denote the size of these sets as $N=|V|$ and $M=|E|$, correspondingly. A \emph{path} is a finite sequence of nodes $\pi = <\! s_{0},s_{1},...,s_{h} \!>$ such that
$s_i \in V$ (for $i \in [0,h]$) and $(s_i, s_{i+1}) \in E$ (for $i \in [0,h-1]$).
Alternatively, a path can be represented by the sequence of its links.
A path is \emph{simple} if all its nodes are distinct.
Given a source node $s \in V$ and a destination node $t \in V$, the \emph{set of all simple paths from $s$ to $t$} is denoted by $P^{(s,t)}$.
Each link $e\in E$ is associated with a failure probability value $p_{e} \in (0,p_{max}]$; we note that these probabilities are often estimated out of the available failure statistics of each network component \cite{Fumagalli:2001}. We assume that each link $e\in E$ fails independently and its failure probability is upper-bounded by some value $p_{max}<1$. Accordingly, we define the \emph{minimum network success probability} as $S_{min}=(1-p_{max})^M$. In addition, each link $e\in E$ is assigned with a positive \emph{weight} $w_{e}$ that represents an additive QoS target such as delay, cost, jitter, etc.

We adopt the \emph{single link failure model},
\footnote{ {Node failures  can also be handled by employing the transformation described in \cite{iqbaldisjoint}, where each node in the network is split into two nodes (say, "in-node" and "out-node") connected by a directed link. Al links that terminate at the original node now terminate at the "in-node" while all links that emerge out of the original node now emanate out of the "out-node". A failure in the original node is captured by a failure of the internal link.}}
which considers handling at most one link failure in the network.  A link is classified as either \emph{faulty} or \emph{operational}: it becomes faulty upon a failure and remains
to be such until it is repaired, otherwise it is operational.
Likewise, we say that a path~$\pi$ is \emph{operational} if it has no faulty link, i.e., for each $e \in \pi$, link~$e$ is operational; otherwise, the path is faulty.

We proceed to formulate the concept of tunable survivability, through the following definitions.

\begin{definition}
\label{def:survivable_connection}
Given a source node $s \in V$ and a destination node $t \in V$, a \emph{survivable connection} is a pair of paths \mbox{$(\pi_{1},\pi_{2}) \in P^{(s,t)} \times P^{(s,t)}$}.
\end{definition}

Survivability is defined as the capability of the network to maintain service continuity in the presence of
failures\TR{ \cite{rfc3386}}. Thus, we say that a survivable connection $(\pi_{1},\pi_{2} )$ is \emph{operational} if either $\pi_{1}$ or $\pi_{2}$ are operational. Under the single link failure model, a survivable connection $(\pi_{1},\pi_{2} )$ is operational \emph{iff} the links that are common to both $\pi_{1}$ and $\pi_{2}$ are operational.
As mentioned, under the single link failure model, a link that is not common to both paths can never cause
a survivable connection to fail; on the other hand, a failure in a common link causes a failure of the entire connection.
Accordingly, as the failure probabilities $\{p_{e}\}$ are independent, we quantify the level of survivability of survivable connections as follows.

\begin{definition}
\label{def:survivability_level}
Given a survivable connection $(\pi_{1},\pi_{2})$ such that $\pi_{1} \cap \pi_{2} \neq \emptyset$, we say that $(\pi_{1},\pi_{2})$ is a \emph{$p$-survivable connection} if $\prod_{e\in \pi_{1} \cap \pi_{2} }(1-p_{e}) \ge p$, i.e., the probability that all common links are operational is at least \emph{$p$}. The value of \emph{$p$} is then termed as the \emph{survivability level} of the connection.
\end{definition}

The above definition formalizes the notion of tunable survivability for the single link failure model. In case that there are no common links between $\pi_{1} $ and $\pi_{2}$, i.e., the paths $\pi_{1} $ and $\pi_{2}$ are disjoint, there is no single failure that can make $(\pi_{1},\pi_{2})$ fail; for this case, $(\pi_{1},\pi_{2} )$ is defined to be a \mbox{\emph{1-survivable connection}}.

In \cite{Banner:2007}, it was shown that, for any network, if there exists a $p$-survivable connection that admits more than two paths, then there exists a $p$-survivable connection that admits exactly two paths. Therefore, we can indeed focus on survivable connections with just two paths.

We proceed to quantify the weight of a survivable connection.
\begin{definition}
\label{def:path_weight}
Given a network $G(V,E)$ and a (non-empty) path $\pi$, its \emph{weight} $W(\pi)$ is defined as the sum of the weight of its links, i.e., \mbox{\(W(\pi) = \sum_{e\in \pi }w_{e}\)}. Accordingly, we define a \emph{weight-shortest path} between two nodes $u,v \in V$ as a path in $G(V,E)$ with minimum weight between $u$ and $v$.
\end{definition}

\comments{
When a connection is composed of two paths, there are several possibilities to define its weight out of the weights of the connection's paths.
A choice of particular interest is to consider  the minimum of the lengths of the two paths.
However, this approach results in strongly NP-complete optimization problems \cite{xu2006complexity}, i.e. even approximate solutions
are computationally intractable.  Alternatively, we can consider the worst (highest) among the weights of the two paths,
yet this also leads to an NP-Hard problem \cite{li1990complexity} (however here tractable approximations are possible, as per below).
Therefore, we adopt the classical approach in the context of connections based on disjoint paths, which attempts to minimize the aggregate weight of the two paths (e.g., \cite{Suurballe:1984}, \cite{4215794}).
Beyond allowing computationally efficient optimal solutions, this approach also provides a $2$-approximation solution to the previous approach, which targets at minimizing the higher length of the
two paths (as shown in \TR{the Appendix }\TON{\cite{TR-QoS-Aware-Survivability}}). However, with tunable survivability, the following refinement is called for.
}

The weight of a $p$-survivable survivable connection $(\pi_{1},\pi_{2} )$ is calculated by the sum of the weights of the links of both paths.
Since a $p$-survivable survivable connection $(\pi_{1},\pi_{2} )$ potentially contains common links, there are
two ways to determine its aggregate weight, namely: counting the weight of a common link
either \emph{once} or \emph{twice}. We shall consider both options, formalized as follows.

\begin{definition}
\label{def:co-weight}
Given a survivable connection  $(\pi_{1},\pi_{2})$, its \emph{CO-weight} $W_{CO}(\pi_{1},\pi_{2})$ is
defined as the sum of its link weights counting the common links \emph{once}, i.e., \mbox{$W_{CO}(\pi_{1},\pi_{2}) = \sum_{e \in \pi_{1} \bigcup \pi_{2}}w_{e}$}.
\end{definition}

\begin{definition}
\label{def:ct-weight}
Given a survivable connection  $(\pi_{1},\pi_{2} )$, its \emph{CT-weight} $W_{CT}(\pi_{1},\pi_{2})$ is defined as the sum of its link weights counting the common links \emph{twice}, i.e., \mbox{$W_{CT}(\pi_{1},\pi_{2}) = \sum_{e \in \pi_{1}} w_{e} + \sum_{e \in \pi_{2}} w_{e}$}.
\end{definition}

The appropriate choice between the two options depends on the QoS metric that the weights  $w_{e}$ represent.
For example, counting the common link once is a good choice for a metric that stands for a monetary cost,
which typically would be paid only once if the link is used by both paths. On the other hand, counting the common link twice is a suitable choice if the QoS metric accounts for an average value
(over the employed paths), e.g. average delay.

For a source-destination pair, there might be several  $p$-survivable connections, among them we would be interested in those that have the best ``quality'', giving rise to several tunable survivability optimization problems.  Each problem, in turn, has its CO-weight ($W_{CO}$) formulation, namely a ``CO-problem'', and its CT-weight ($W_{CT}$) formulation, namely a ``CT-problem''.
\comments{
The following definition formalizes one of the problems.

\begin{definition}
\label{def:CT-CQMS}
\textbf{\emph{CT-Constrained QoS Max-Survivability (CT-CQMS) Problem}}:
Given are a network $G(V,E)$, a source node $s \in V$, a destination node $t \in V$ and a QoS bound~${B}$. Find a survivable connection \mbox{$(\pi_{1},\pi_{2}) \in P^{(s,t)} \times P^{(s,t)}$} from $s$ to $t$ such that:
\[
\max{\prod_{e \in \pi_{1} \cap \pi_{2}} (1-p_{e})}
\]
\[
s.t.\,\, W_{CT}(\pi_{1},\pi_{2})  \leq {B}.
\]
\end{definition}

The CO version of the above problem, namely the \emph{CO-Constrained QoS Max-Survivability} (CO-CQMS) Problem, is
defined in the same way but replacing the term $W_{CT}(\pi_{1},\pi_{2})$ with the term $W_{CO}(\pi_{1},\pi_{2})$.
In \cite{TR-QoS-Aware-Survivability}, we also consider dual problems, where we try to minimize the connection's weight while observing a constraint on the level of survivability. Their solution is obtained quite simply out of the solutions of the problems defined above.

In the following sections, we shall establish algorithmic solutions for the above problems, namely CT-CQMS and CO-CQMS.
We begin by establishing some interesting structural properties of the CT-problem.
}
{
The following definitions formalize the different versions of the problem.

\begin{definition}
\label{def:CT-CQMS}
\textbf{\emph{CT-Constrained QoS Max-Survivability (CT-CQMS) Problem}}:
Given are a network $G(V,E)$, a source node $s \in V$, a destination node $t \in V$ and a QoS bound~${B}$. Find a survivable connection \mbox{$(\pi_{1},\pi_{2}) \in P^{(s,t)} \times P^{(s,t)}$} from $s$ to $t$ such that:
\[
\max{\prod_{e \in \pi_{1} \cap \pi_{2}} (1-p_{e})}
\]
\[
s.t.\,\, W_{CT}(\pi_{1},\pi_{2})  \leq {B}.
\]
\end{definition}

\begin{definition}
\label{def:CT-CSMQ}
\textbf{\emph{CT-Constrained Survivability Min-QoS (CT-CSMQ) Problem}}:
Given are a network $G(V,E)$, a source node $s \in V$, a destination node $t \in V$ and a survivability level~${S} \in [S_{min},1]$. Find a survivable connection \mbox{$(\pi_{1},\pi_{2}) \in P^{(s,t)} \times P^{(s,t)}$} from $s$ to $t$ such that:
\[
\min\,\, { W_{CT}(\pi_{1},\pi_{2})}
\]
\[
s.t.\,\, \prod_{e \in \pi_{1} \cap \pi_{2}} (1-p_{e}) \geq {S}.
\]
\end{definition}
\comments{
\begin{definition}
\label{def:CO-CQMS}
\textbf{\emph{CO-Constrained QoS Max-Survivability (CO-CQMS) Problem}}:
Given are a network $G(V,E)$, a source node $s \in V$, a destination node $t \in V$ and a QoS bound~${B}$. Find a survivable connection \mbox{$(\pi_{1},\pi_{2}) \in P^{(s,t)} \times P^{(s,t)}$} from $s$ to $t$ such that:
\[
\max\ {\prod_{e \in \pi_{1} \cap \pi_{2}} (1-p_{e}) }
\]
\[
s.t.\,\, W_{CO}(\pi_{1},\pi_{2})  \leq {B}.
\]
\end{definition}

\begin{definition}
\label{def:CO-CSMQ}
\textbf{\emph{CO-Constrained Survivability Min-QoS (CO-CSMQ) Problem}}:
Given are a network $G(V,E)$, a source node $s \in V$, a destination node $t \in V$ and a survivability level~${S} \in [S_{min},1]$. Find a survivable connection \mbox{$(\pi_{1},\pi_{2}) \in P^{(s,t)} \times P^{(s,t)}$} from $s$ to $t$ such that:
\[
\min\,\, { W_{CO}(\pi_{1},\pi_{2})}
\]
\[
s.t.\,\, \prod_{e \in \pi_{1} \cap \pi_{2}} (1-p_{e} ) \geq {S}.
\]

\end{definition}
}

The CO version of the above problems, namely the \emph{CO-Constrained QoS Max-Survivability} (CO-CQMS) Problem  and \emph{CO-Constrained Survivability Min-QoS} (CO-CSMQ) Problem, are
defined in the same way but replacing the term $W_{CT}(\pi_{1},\pi_{2})$ with the term $W_{CO}(\pi_{1},\pi_{2})$.

In the following sections, we will establish algorithmic solutions for the defined problems. We begin by establishing some interesting structural properties of CT-problems.
}

\section{The Structure of CT Solutions}
\label{sec:ct_charc}

As explained, CT-problems are an important class in which the QoS metric represents either an average or aggregate measure over the employed survivable connection, e.g., the average delay over the two paths of the connection.
We proceed to show that, when addressing the optimization problems of the CT class, the links that may affect the survivability level of the optimal solution are restricted to a (typically small) subset of the network's links. We start with the following definitions.

\begin{definition}
\label{def:critical_link}
Given a survivable connection  $(\pi_{1},\pi_{2})$, a \emph{critical link} is a link $e \in E$ that is common to both paths $\pi_{1}$ and $\pi_{2}$. Accordingly, the set of critical links of a survivable connection is defined as $ \mathbbm{C} (\pi_{1},\pi_{2})= \{ e | e \in \pi_{1} \cap \pi_{2}\}$.
\end{definition}

\begin{definition}
\label{def:weight-shortest-paths-set}
Given a source $s$ and a destination $t$, $L^{(s,t)}$ is the set of all the weight-shortest paths between $s$ and $t$. Note that $L^{(s,t)}~\subseteq~P^{(s,t)}$.
\end{definition}

\begin{definition}
\label{def:in-all-weight-shortest-paths_link}
Given a source node $s \in V$ and a destination node $t \in V$, an \emph{in-all-weight-shortest-paths link}  is a link $e \in E$ that is common to all paths in $L^{(s,t)}$. Accordingly, the set of in-all-weight-shortest-paths links is defined as  $\mathbbm{L}~=~\{e |~e\in~\bigcap_{\pi \in L^{(s,t)}}\pi\}$.
\end{definition}

Note that if there is a unique weight-shortest path between $s$ and $t$, i.e. $|L^{(s,t)}|=1$, then $\mathbbm{L}$ precisely consists of its links.
Moreover, $\mathbbm{L}$ is a subset of the set of links of any weight-shortest path.
We are ready to present the main result of this section.
\begin{theorem}
\label{allin_theorem}
For \emph{any bound} ${B}$ on the additive end-to-end QoS, {a} (any) survivable connection $(\pi_{1},\pi_{2})$ that is an optimal solution of the respective CT-Constrained QoS Max-Survivability Problem (per Def. \ref{def:CT-CQMS}) is such that \textbf{\emph{all its critical links are in-all-weight-shortest-paths links}}. That is,  $\mathbbm{C}(\pi_{1},\pi_{2}) \subseteq \mathbbm{L} $.
\end{theorem}

\begin{IEEEproof}
Let $(\pi_{1},\pi_{2})$ be  an optimal survivable connection that solves the CT-CQMS Problem (Def. \ref{def:CT-CQMS}). Assume by contradiction that there is a critical link $e \in \pi_{1} \cap \pi_{2} $ that is not an \emph{in-all-weight-shortest-paths link}, i.e,  $ \exists e \,|\, e\in \pi_{1} \cap \pi_{2} \, \wedge \, e \notin \bigcap_{\pi \in L^{(s,t)}}\pi\  $. Let  $v_{i}$ and $v_{j}$  be the nodes of this critical link $e$, henceforth denoted as $ v_{i} \to v_{j} $. From the assumption, there is a weight-shortest path $\pi_{short}$ that does not contain the critical link $v_{i} \to v_{j}$, i.e., $v_{i} \to v_{j} \notin \pi_{short}$. Moreover, $\pi_{short}$ is not identical to $\pi_{1}$ nor $\pi_{2}$, since $\pi_{short}$ does not contain $ v_{i} \to v_{j} $, a common link of both $\pi_{1}$ and $\pi_{2}$.
Consider all nodes that are common to $\pi_{short}$ and to at least one of the optimal survivable connection paths, $\pi_{1}$ or $\pi_{2}$. Denote by $v_{a}$ the last such common node on the corresponding sub-path from the source $s$ to $v_{i}$. Similarly, $v_{p}$ denotes the first such common node on the corresponding sub-path from $v_{j}$ to the destination $t$.  Note that $v_{a}$, $v_{p}$ can include  $v_{i}$, $v_{j}$ as well as the source $s$ and the destination $t$, respectively.
From the assumption, a pair of disjoint paths between $v_{a}$ to $v_{p}$ necessarily exists.
Moreover, $\pi_{short}$ intersects with either $\pi_{2}$ or $\pi_{1}$, only in the sub-paths between  $s$  to  $v_{a}$ or $v_{p}$ to $t$.
 Through Fig. \ref{fig:intersection}, we proceed to consider two possible cases of an intersection between $\pi_{short}$ (the full-lines path) and the optimal survivable connection $(\pi_{1},\pi_{2})$ (the dashed-lines paths).

\begin{figure}
    \centering

    \begin{subfigure}[tb]{0.4\textwidth}
        \centering
        \includegraphics[width=\textwidth]{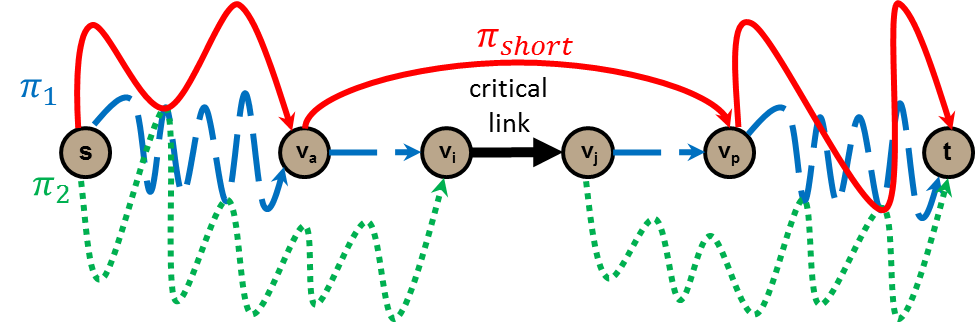}
        \caption{Nodes $v_{a}$ and $v_{p}$ belong to the same path }
\label{fig:same_path}
    \end{subfigure}%

    \begin{subfigure}[tb]{0.4\textwidth}
        \centering
        \includegraphics[width=\textwidth]{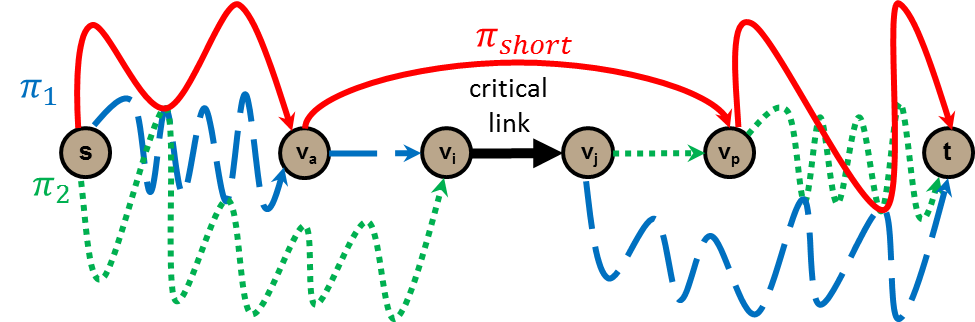}
         \caption{Nodes $v_{a}$ and $v_{p}$ belong to different paths }
  \label{fig:different_path}
    \end{subfigure}%

  \caption{Intersection between $\pi_{short}$ and the optimal survivable connection $(\pi_{1},\pi_{2})$ }
  \label{fig:intersection}
\end{figure}

In the first case, illustrated in Fig. \ref{fig:same_path}, $v_{a}$ and $v_{p}$ belong to the same path in the optimal survivable connection $(\pi_{1},\pi_{2})$. Without loss of generality,
we assume that $v_{a}, v_{p} \in \pi_{1}$. Consider the pair of sub-paths from $v_{a}$ to $v_{p}$, where one path
contains  links from $\pi_{1}$  and the other path contains  links from $\pi_{short}$  denoted as $(\pi_{1}^{sub}, \pi_{short}^{sub})$. It is obvious from the definition that  $(\pi_{1}^{sub}, \pi_{short}^{sub})$ are disjoint. Denote by $\pi_{1}^{pre}$ the sub-path of $\pi_{1} $ from the source $s$ to $v_{a}$, and  by $\pi_{1}^{post}$ the sub-path of $\pi_{1}$  from $v_{p}$ to the destination $t$. Now, define a new path $\pi_{new}$ described as $\pi_{1}^{pre} \, \to v_{a} \to \pi_{short}^{sub} \to v_{p} \to \pi_{1}^{post}$, composed by the sub-paths $\pi_{1}^{pre}$, $\pi_{short}^{sub}$  and $\pi_{1}^{post}$. Consider the survivable connection $(\pi_{new},\pi_{2} )$. Since $\pi_{new} \cap \pi_{2}$ does not include the critical link ${v_{i}\to v_{j}}$, the survivability level of $(\pi_{new},\pi_{2})$ is higher than that of $(\pi_{1},\pi_{2})$, i.e.
$$\prod_{e \in \pi_{new} \cap \pi_{2} }(1-p_{e}) > \prod_{e \in \pi_{1} \cap \pi_{2}}(1-p_{e} ).$$
Also, since for additive metrics a sub-path of a weight-shortest path is also a weight-shortest path between its endpoints, we have that $W(\pi_{short}^{sub}) \le W(\pi_{1}^{sub})$.
Therefore, the CT-weight of $(\pi_{new},\pi_{2})$ is not larger than that of $(\pi_{1},\pi_{2})$,
i.e. $W_{CT}(\pi_{new},\pi_{2} ) \le W_{CT}(\pi_{1},\pi_{2})$.
Thus, the survivable connection $(\pi_{new},\pi_{2} )$  strictly outperforms $(\pi_{1},\pi_{2})$ in terms of survivability while not incuring a higher weight, which contradicts the assumption that $(\pi_{1},\pi_{2})$ is optimal.

In the second case, illustrated in Fig. \ref{fig:different_path},  $v_{a}$ and $v_{p}$ belong to different paths in the optimal survivable connection $(\pi_{1},\pi_{2})$. Without loss of generality, we assume that $v_{a} \, \in \pi_{1}$ and $v_{p} \in \pi_{2} $.
Denote  $\pi_{1}$'s sub-paths from the source to
$v_{i}$  and from $v_{j} \,$ to the destination as  $\pi_{1}^{pre}$ and  $\pi_{1}^{post}$, respectively. Similarly, we define  $\pi_{2}^{pre}$ and  $\pi_{2}^{post}$.
Now, consider the survivable connection $(\gamma_{1},\gamma_{2})$ described as $(\pi_{1}^{pre} \to v_{i} \to v_{j} \to \pi_{2}^{post} \,,\,
\pi_{2}^{pre} \to v_{i} \to v_{j} \to \pi_{1} ^{post} )$, composed by the sub-paths $\pi_{1}^{pre}$, $\pi_{1}^{post}$, $\pi_{2}^{pre}$ and  $\pi_{2}^{post}$ and the critical link $ v_{i} \to v_{j} $. It has precisely the same links as $(\pi_{1},\pi_{2} )$,
hence it has the \emph{same survivability level}, i.e.
$\prod_{e\in \pi_{1} \cap \pi_{2} }(1-p_{e} ) =\prod_{e\in \gamma_{1} \cap \gamma_{2} }(1-p_{e})$, and the same CT-weight, i.e.
$W_{CT}(\pi_{1},\pi_{2} ) = W_{CT}(\gamma_{1},\gamma_{2})$.
Furthermore, $v_{a}$ and $v_{p}$ belong to the same path in the survivable connection $(\gamma_{1},\gamma_{2} )$ i.e., we are back
in the realm of the first case.
\end{IEEEproof}

\TR{
\begin{corollary}
\label{all_in_corollary}
For \emph{any bound} ${S}$ on the survivability level, there is a survivable connection  $(\pi_{1},\pi_{2})$ that is an optimal solution of the respective CT-Constrained Survivability Min-QoS (CT-CSMQ) Problem (Def. \ref{def:CT-CSMQ}) such that \textbf{\emph{all its critical links are in-all-weight-shortest-paths links}}. That is,  $\mathbbm{C}(\pi_{1},\pi_{2}) \subseteq \mathbbm{L} $.
\end{corollary}

\begin{IEEEproof}
Consider a CT-CSMQ Problem for some bound ${S}$ on the survivability level.
Let ${B}$ be the minimum CT-weight of \emph{any} optimal survivable connection solution of the above problem instance.
On the same network and for the same source and destination nodes, consider the CT-CQMS Problem with ${B}$ as the bound on the additive end-to-end QoS, and denote by $(\pi_{1},\pi_{2})$ an optimal solution for this problem
and by $\mathcal{S^*}$ and $\mathcal{B^*}$ its survivability level and CT-weight, respectivily. Clearly,  $\mathcal{S^*}\geq{S}$, and, moreover, $\mathcal{B^*}=B$.
Thus, $(\pi_{1},\pi_{2})$ is also an optimal feasible solution to the original CT-CSMQ Problem with bound ${S}$ on the survivability level. According to Theorem \ref{allin_theorem}, all the critical links of $(\pi_{1},\pi_{2})$ are in-all-weight-shortest-paths links.
\end{IEEEproof}
}

We shall employ the above property of the CT-problems in order to reduce the computational complexity of the solution algorithms. Furthermore, we shall exploit this property in order to establish a design scheme for efficiently upgrading the performance of the network in terms of survivability.

\section{Establishing QoS Aware $p$-Survivable Connections is NP-Hard}
\label{sec:Intractability}
{In this section we will prove that our optimization problems, namely CT-CQMS, CT-CSMQ, CO-CQMS, CO-CSMQ  are \emph{NP-Hard}.}
\comments{In this section we will prove that our optimization problem, namely CT-CQMS, is \emph{NP-Hard}.}
 We provide a reduction from the well-known, NP-Complete, \emph{Partition Problem} (PP) \cite{Garey:1979}.

The Partition Problem is defined as follows: Given are a finite set $A$ and a size $s(a) \in \mathbb{Z}^{+}$ for each $a \in A$; is there a subset $A' \subseteq A$ such that $\sum_{a \in A'}s(a) = \sum_{a \in A \setminus A'}s(a)$?

{Both of our optimization problems, namely CT-CQMS and CT-CSMQ,}
\comments{The CT-CQMS problem}
can be reduced to the following decision problem, denoted as the {\emph{CT-Restricted Weight Survivability Connection} (RWSC)} problem. Given are a source node $s \in V$, a destination node $t \in V$, a QoS bound~${B}$ and a survivability level~${S}$. Is there a survivable connection \mbox{$(\pi_{1},\pi_{2}) \in P^{(s,t)} \times P^{(s,t)}$} from $s$ to $t$ in which $\prod_{e \in \pi_{1} \cap \pi_{2} }(1-p_{e} ) \geq {S}$ and $W_{CT}(\pi_{1},\pi_{2})  \leq {B}$?

\begin{theorem}
\label{theorem:NP-hard}
Problem CT-RWSC is NP-Complete.
\end{theorem}

\begin{figure}[tb]
  \centering
    \includegraphics[width=0.48\textwidth]{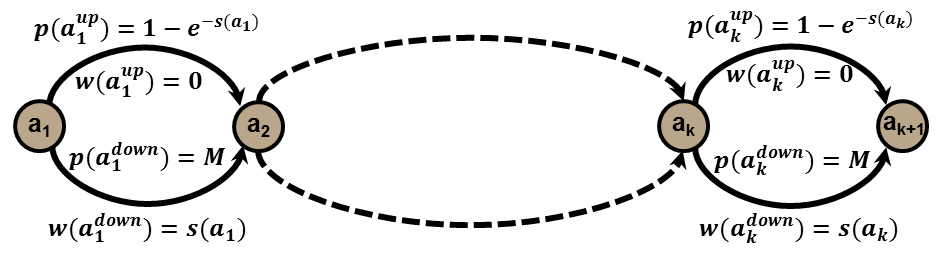}
  \caption{RWSC graph constructed from the Partition Problem}
\label{fig:intract}
\end{figure}

\begin{IEEEproof}
Clearly, CT-RWSC is in NP, since for a given  survivable connection  $(\pi_1,\pi_2)$, we can polynomially check whether $\prod_{e \in \pi_{1} \cap \pi_{2} }(1-p_{e} ) \geq {S}$ and $W_{CT}(\pi_{1},\pi_{2})  \leq {B}$ by calculating these two metrics and checking the links of the survivable connection.

Through the following reduction from Problem PP, i.e. $PP \leq_{p} CT-RWSC$. Consider an instance of
Problem PP, i.e., a set $A=\{a_{1},...,a_{i},...,a_{k}\}$ where $\sum_{a_{i} \in A} s(a_{i}) = T$. Construct the direct graph illustrated in Fig. \ref{fig:intract}, as follows. Create a node $a_i$ for each element in the set $A$ and an additional node $a_{k+1}$. Connect every two adjacent nodes, $a_i$ and $a_{i+1}$, by a pair of links: a top link $a_i^{up}$ and a bottom link $a_i^{down}$. Determine node $a_1$ as the source and node $a_{k+1}$ as the destination. Note that each node in the constructed graph, except the source and the destination, has in-degree and out-degree equal to 2.  The top and bottom link weight and failure probability values are set according to the equivalent element size $s(a_{i})$, as follows. Set $a_i^{up}$ weight as $w(a_i^{up})=0$ and $a_i^{up}$ failure probability as $p(a_i^{up})=1 - e^{-s(a_i)}$. Note that $0 < p(a_i^{up}) < 1$ due to the negative exponent. Set $a_i^{down}$ weight as $w(a_i^{down})=s(a_i)$ and $a_i^{down}$ failure probability as $p(a_i^{down}) = M$, where $p_{max}< M <1$. We remind that $p_{max}$ is an upper-bound for the failure probability.
Consider the CT-RWSC problem for the above graph where {the upper bound ${B}$ is set to $\frac{T}{2}$ and the lower bound ${S}$  is set to  $e^{\frac{-T}{2}}$}. Our proof is based on the following two lemmas.
\end{IEEEproof}

\begin{lemma}
If there is no solution to the given CT-RWSC problem, there is no solution to the PP problem.
\end{lemma}
\begin{IEEEproof}
Assume by contradiction that there is a solution to the PP problem, denoted by $A'$ where $\sum_{a \in A'}s(a) = \frac{T}{2}$. Consider a survivable connection $(\pi_{1},\pi_{2})$  in the constructed graph (Fig. \ref{fig:intract}) where its common links are the $a_i^{up}$ links associated with $a_i \in A'$.  Accordingly, if $a_i \in A \setminus A'$ then both $a_i^{up}$ and $a_i^{down}$ links belongs to the survivable connection $(\pi_{1},\pi_{2})$.
Therefore, the survivability level of $(\pi_{1},\pi_{2})$ is  $\prod_{e \in \pi_{1} \cap \pi_{2}} (1-p_{e}) = \prod_{a_i \in A'}e^{-s(a_i)} = e^{\frac{-T}{2}}$.  Since top links have zero weight, $w(a_i^{up})=0$, they do not contribute to the total survivable connection weight. Hence,  the weight of  $(\pi_{1},\pi_{2})$ is the sum of the bottom links weight, $w(a_i^{down})$, in the survivable connection, i.e. $W_{CT}(\pi_{1},\pi_{2}) = \sum_{a_i \in A \setminus A'}s(a_i)=\frac{T}{2}$. Thus, $(\pi_{1},\pi_{2})$  is a solution to the given CT-RWSC problem, which is a contradiction.
\end{IEEEproof}

\begin{lemma}
If there is a solution $(\pi_{1},\pi_{2})$ to the given CT-RWSC problem, there is a solution to the PP problem.
\end{lemma}

\begin{IEEEproof}
Clearly, a bottom link $a_i^{down}$ will never be a common link of the solution $(\pi_{1},\pi_{2})$, due to the high failure probability set to this link, as defined $p_{max} < M <1$.
Thus, for each pair of links, $a_i^{up}$ and $a_i^{down}$, connecting two nodes, $a_i$ and $a_{i+1}$, the solution $(\pi_{1},\pi_{2})$ may consist of either top link $a_i^{up}$, common to the two paths, or the  pair of links, $a_i^{up}$ and $a_i^{down}$, each assigned to one of the paths.
Note that only the common top links contribute  to the connection's
survivability level, and only the bottom links contribute the connection's total weight.
{Denote by $S_{1}$ the set of elements that are associated with common top links  of $(\pi_{1},\pi_{2})$ and by $S_{2}$ the set of elements that are associated with the other links of $(\pi_{1},\pi_{2})$.
Note that $S_{1}$ and $S_{2}$ are complementary subsets of $A$.
Therefore, for a given solution $(\pi_{1},\pi_{2})$,  $\prod_{e \in \pi_{1} \cap \pi_{2}} (1-p_{e}) \leq e^{\frac{-T}{2}}$ (resp. $\prod_{e \in \pi_{1} \cap \pi_{2}} (1-p_{e}) \geq e^{\frac{-T}{2}}$) \emph{iff}
$\sum_{a \in S_1}s(a) \geq \frac{T}{2}$ (resp. $\sum_{a \in S_1}s(a) \leq \frac{T}{2}$)  \emph{iff}
$\sum_{a \in S_2}s(a) \leq \frac{T}{2}$ (resp. $\sum_{a \in S_2}s(a) \geq \frac{T}{2}$)  \emph{iff}
$W_{CT}(\pi_{1},\pi_{2}) \leq \frac{T}{2}$ (resp. $W_{CT}(\pi_{1},\pi_{2}) \geq \frac{T}{2}$).
Since $(\pi_{1},\pi_{2})$ should meet both bounds, we have $\prod_{e \in \pi_{1} \cap \pi_{2}} (1-p_{e}) = e^{\frac{-T}{2}}$ and $W_{CT}(\pi_{1},\pi_{2}) = \frac{T}{2}$.
Thus, we identified a subset $S_1 \subseteq A$ such that $\sum_{a_i \in S_1}s(a_i) = \sum_{a_i \in A \setminus S_1}s(a_i)= \frac{T}{2}$.}
\end{IEEEproof}

\TON{A similar reduction presented in \cite{TR-QoS-Aware-Survivability} shows that the CO-CQMS problem is also NP-hard.}

\TR{
Both of our optimization problems, namely CO-CQMS and CO-CSMQ, can be reduced to the following decision problem, denoted as the \emph{CO-RWSC} problem.
Given are a source node $s \in V$, a destination node $t \in V$, a QoS bound~${B}$ and a survivability level~${S}$. Is there a survivable connection \mbox{$(\pi_{1},\pi_{2}) \in P^{(s,t)} \times P^{(s,t)}$} from $s$ to $t$ in which
$\prod_{e \in \pi_{1} \cap \pi_{2} }(1-p_{e} ) \geq {S}$ and $W_{CO}(\pi_{1},\pi_{2})  \leq {B}$?

\begin{corollary}
    The CO-RWSC problem is NP-Hard.
\end{corollary}
\begin{IEEEproof}
As defined in section \ref{sec:formulation}, the difference between the CO and CT problems is based on the effect of the common links weights of a survivable connection  in the total solution weight.
Previously, in the proof of Theorem \ref{theorem:NP-hard}, we presented a construction (Fig. \ref{fig:intract}) in which the weight of the top link is set to 0, i.e. $w(a_i^{up})=0$. As mentioned, in this construction, a bottom link $a_i^{down}$ will never be a common link of the solution $(\pi_{1},\pi_{2})$, since the common links of any optimal survivable connection solution are composed only of top links $a_i^{up}$.
Therefore, the common links of the presented construction  did not affect the total solution weight. Thus, the previous proof of Theorem \ref{theorem:NP-hard} can be applied also to CO problems.
\end{IEEEproof}
}

Nonetheless, the Partition Problem is a weakly NP-complete problem and admits pseudo-polynomial time algorithms and approximation schemes \cite{Garey:1979}. In fact, the problem has been called "The Easiest Hard Problem" \cite{hayes2002easiest}. Indeed, we proceed to establish pseudo-polynomial time algorithms and approximation schemes for our optimization problems.

\section{Establishing QoS Aware $p$-Survivable Connections}
\label{sec:algo}
In the previous section, we prove that our previously formulated optimization problems,
{namely CT-CQMS, CT-CSMQ, CO-CQMS and CO-CSMQ,}
\comments{namely CT-CQMS and CO-CQMS,}
are \emph{NP-Hard}.
However, efficient solution schemes are still possible. Indeed, in this section, we shall establish  exact solutions of pseudo-polynomial complexity, and near (i.e., $\epsilon$-optimal) solutions of polynomial complexity, for the considered problems.

The solution approach is based on a graph transformation  that reduces our problem to a standard \emph{Restricted Shortest Path} (RSP) problem. We recall that RSP is the problem of finding a
shortest (in terms of an additive metric) path while obeying an additional (additive) constraint, as follows.
\begin{definition}
\label{def:RSP}
\textbf{\emph{Restricted Shortest Path (RSP) Problem}}:
Given is a network $G(V,E)$ where each link $e\in E$ is associated with a length $l_e$ and a time $t_e$. Let $T$ be a positive integer and $s,t \in V$ be the source and the destination nodes, respectively. Find a path \mbox{$\pi$} from $s$ to $t$ such that:
\[
\min\,\, { \sum_{e \in \pi}l_{e}}\ \ \
s.t.\,\, { \sum_{e \in \pi}t_{e}} \leq {T}.
\]
\end{definition}
Although the RSP problem is known to be
NP-Hard \cite{Garey:1979}, the literature provides several pseudo-polynomial solutions\TR{ \cite{lawler:2001}} \cite{joksch:1966} as
well as $\epsilon$-optimal Fully Polynomial Time Approximation Schemes (FPTAS) \cite{Hassin:1992} \cite{Lorenz:2001213},
which we employ in order to solve our problems.
Moreover, we use the findings of Section \ref{sec:ct_charc} in order to further reduce the complexity of the solutions for  the \emph{CT} problem.

\subsection{Pseudo-Polynomial Schemes for Establishing CO-QoS Aware $p$-Survivable Connections}
\label{sec:algo-CO-QoS}

We begin by establishing  a pseudo-polynomial algorithmic scheme for solving
\comments{the CO-CQMS Problem, denoted as the \emph{CO-QoS Aware Max Survivable Connection} (CO-QAMSC) Algorithm and specified in Fig. \ref{algo:CO-CT}. Note that the algorithm does not include the dashed-boxed text (with gray background), which shall be later used for handling the CT-CQMS Problem.}
the two CO problems, namely the CO-CSMQ Problem  and the CO-CQMS Problem.
The method employs two well-known algorithms:
the first, \emph{Edge-Disjoint Shortest Pair} (EDSP) Algorithm \cite{Suurballe:1984}, finds two edge-disjoint paths with minimum sum of edge weight between two nodes in a weighted directed graph; the second is a pseudo-polynomial algorithmic scheme, such as \cite{joksch:1966}, for solving the NP-Hard RSP problem.
We proceed to present an algorithmic scheme for solving the both CO-Problems specified in Fig. \ref{algo:CO-CT}. We denote the algorithmic schemes as the \emph{CO-Tunable Survivable Connection Min-QoS} (CO-TSCMQ) Algorithm, for the CO-CSMQ Problem,   and  the \emph{CO-QoS Aware Max Survivable connection} (CO-QAMSC) Algorithm, for the CO-CQMS Problem. Note that the algorithm does not include the dashed-boxed text (with gray background), which shall be later used for handling the CT-QoS Aware Problems. Moreover, the CO-TSCMQ and  the CO-QAMSC algorithms differ from each other only by the text in Stage 2 inside the double-framed-box and the dashed-box, respectively. The scheme consists of the three following stages.

{
\begin{algorithm}
\DontPrintSemicolon
\SetAlgoNoLine
\Indm
\KwIn{
$G(V,E)$-network,
$s$-source,
$t$-destination,
${p_{e}}$-link failure probability,
${w_{e}}$-link weight,
${B}$-weight constraint}
\KwVar{
$\tilde{G}(\tilde{V},\tilde{E})$-transformed network,
$\tilde{s}$-source,
$\tilde{t}$-destination,
${l_{\tilde{e}}}$-link {length},
${t_{\tilde{e}}}$-link {time},
$\pi_{min}$ -a weight-shortest path,
$\tilde{\pi}$ -RSP solution
}
\raisebox{-4ex}{
\colorbox{light-gray}{
\dashbox{4}(250,40)
{$
\begin{array}{l}
{\KwStage\,0\,-\,Shortest\,Path\,Search} \\
{\textbf{0.1)}\,Employ\,a\,\emph{Shortest-Path Algorithm}\cite{Tarjan:1984}\,between\,s\,and\,t}\\
{in\,order\,to\,find\,a\,weight\,shortest\,path\,\pi_{min}}.
\end{array}
$}
}
}
\comments{
\colorbox{light-gray}{\KwStage{0 - Shortest-Path Search}}\;
\Indp
\colorbox{light-gray}{\textbf{0.1)} Employ a \emph{Shortest-Path Algorithm}\cite{Tarjan:1984} between $s$ and $t$}
\colorbox{light-gray}{in order to find a weight-shortest path $\pi_{min}$.}
\Indm
}

\KwStage{1 - Transformed network $\tilde{G}(\tilde{V},\tilde{E})$ construction}\;
\Step{1.1)} $\tilde{V} \shortleftarrow V$.\;
\Step{1.2)} \ForEach {$e:u\shortrightarrow v \in \framebox{E}$ \MyColorBox{\dashbox{4}(22,10){$\pi_{min}$}}}{
  - Construct a ``simple link'' $\tilde{e}$ between $\tilde{u}$ and $\tilde{v}$.\;
  - Assign $l_{\tilde{e}}$ to be \framebox{$w_{e}$} \MyColorBox{\dashbox{4}(25,10){$2 \cdot w_{e}$}}.\;
  - Assign $t_{\tilde{e}}$ to be $-ln(1-p_{e})$.\;
}	
\Step{1.3)} \ForEach {$u \in \framebox{V}$ \MyColorBox{\dashbox{4}(22,10){$\pi_{min}$}}}{
    \ForEach {$v \in  \framebox{V}$ \MyColorBox{\dashbox{4}(22,10){$\pi_{min}$}}}{

- Employ the \emph{EDSP Algorithm} \cite{Suurballe:1984} between $u$ and $v$.\;
\If{ (there is a solution to the EDSP Algorithm)} {
- Construct a ``disjoint link'' $\tilde{e}$ between $\tilde{u}$ and $\tilde{v}$.\;
- Assign $l_{\tilde{e}}$ to be the sum of $w_{e}$'s of the
Edge-Disjoint Shortest Pair of Paths (EDSPoP).\;
- Assign $t_{\tilde{e}}$ to be 0.\;
}
}
}
\KwStage{2 - RSP Calculation}

\Step{2.1)} Find a restricted shortest path $\tilde{\pi}$ of the RSP problem (Def. \ref{def:RSP}) by employing \cite{joksch:1966} in the transformed network $\tilde{G}(\tilde{V},\tilde{E})$ where
\doublebox{$T=B$} \raisebox{-0.8ex}{\dashbox{4}(50,13){$T=-\ln{S}$}}.


\Step{2.2)} \If{ there is no feasible solution for the RSP }
{ - \Return {Fail}}

\KwStage{3- Survivable Connection $(\pi_1, \pi_2)$ Construction}

\Step{3.1)} \ForEach {$\tilde{e} \in \tilde{\pi}$}{
		\If {$\tilde{e}$ is a ``simple link''}
		{
			- $\pi_1$ and $\pi_2$ contain the corresponding $e$ link.\;
		}
		\ElseIf {$\tilde{e}$ is a ``disjoint link''}
		{
			- $\pi_1$ contains the links of one the paths of the EDSPoP represented by $\tilde{e}$.\;
			- $\pi_2$ contains the links of the other path of the EDSPoP represented by $\tilde{e}$.\;
		}
}
return $(\pi_1, \pi_2)$
\caption{CO/CT-QoS Aware Max Survivable Connection/ \\Tunable Survivable Connection Min-QoS \\ (CO/CT-QAMSC/TSCMQ) Algorithm}
\label{algo:CO-CT}
\end{algorithm}
}

\begin{figure}
    \centering
    \begin{subfigure}[tb]{0.4\textwidth}
        \centering
        {\includegraphics[width=\textwidth]{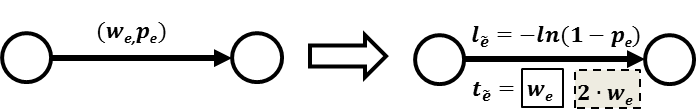}}
        \caption{simple links}
        \label{fig:simple_tranformation}
    \end{subfigure}%

    \begin{subfigure}[tb]{0.4\textwidth}
        \centering
         \includegraphics[trim = 0mm 0mm 0mm 17mm, clip, width=\textwidth]{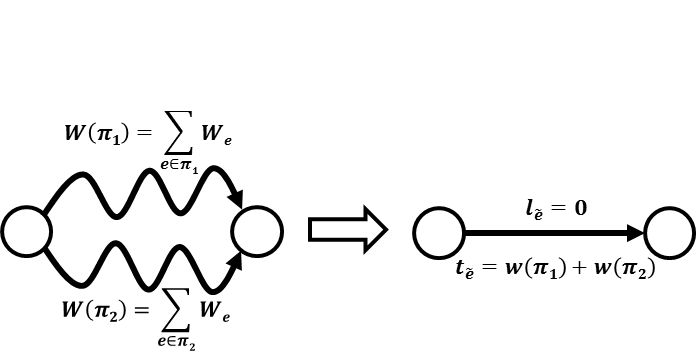}
        \caption{disjoint links}
        \label{fig:disjoint_tranformation}
    \end{subfigure}%
  \caption{CO\TON{/CT} links transformation}
\label{fig:algo_tranformation}
\end{figure}

The first stage comprises the construction of a transformed network $\tilde{G}(\tilde{V},\tilde{E})$ {constituting an input for an RSP algorithm in the next stage, where each link is associated with two metrics: a length $l_{\tilde{e}}$ and a time $t_{\tilde{e}}$}.
Specifically, the transformed network consists of two types of links, as follows. The first one, denoted as \emph{simple link}, consists of the original network links. The {length} of a simple link is set to be the weight of the original link, i.e. $t_{\tilde{e}}=w_{e}$. The {time} of a simple link is set to
$t_{\tilde{e}} = -ln(1-p_{e})$, thus transforming our multiplicative (survivability) metric into an additive one.
The second type of links, denoted as \emph{disjoint link}, consists of additional links representing possible
\emph{Edge-Disjoint Shortest Pair of Paths} (EDSPoP) between pairs of nodes in the network. The {length} of a disjoint link is set to be the weight of the EDSPoP between these two nodes, which we compute by employing  the EDSP Algorithm \TR{\cite{Bhandari:1999} }\cite{Suurballe:1984}. The {time} of a disjoint link is set to be $0$, due to the fact that a disjoint path provides full protection against a single link failure. Fig. \ref{fig:algo_tranformation} illustrates these transformations{, where the dashed-boxed text (with gray background) should be disregarded at this stage}.
Given the above transformed network $\tilde{G}(\tilde{V},\tilde{E})$, the second stage calculates a restricted shortest path according to the desired version of the algorithm distinguished by the framed-box, where the CO-QAMSC algorithm is marked by the double-framed-box  and the CO-TSCMQ algorithm is marked by the dashed-box.
{We remind that the CO-QAMSC algorithm aims to find a survivable connection with maximum survivability level  upper bounded by a QoS constraint $B$, thus the parameters of the RSP problem (Def. \ref{def:RSP}) are set to $l_e=l_{\tilde{e}}$, $t_e=t_{\tilde{e}}$ and $T=B$. In contrast, the CO-TSCMQ algorithm aims to find a survivable connection with minimum  CO-weight  lower bounded by a survivability level constraint $S$, thus the parameters of the RSP problem (Def. \ref{def:RSP}) are set to $l_e=t_{\tilde{e}}$, $t_e=l_{\tilde{e}}$ and $T=-\ln{S}$.}
Note that, for the CO-TSCMQ algorithm, the constraint expression $T=-\ln{S}$ usually is a non-integer number, therefore the pseudo-polynomial algorithm  depends on the precision the expression.
Specifically, the CO-QAMSC algorithm finds a pair of paths that minimizes
\[ -\sum_{\substack{e \in \pi_{1} \cap \pi_{2} }} ln(1-p_{e}) = -ln \prod_{\substack{ e \in \pi_{1} \cap \pi_{2} }} (1-p_{e}) \]
and, therefore, maximizes \( \prod_{ \substack{ e \in \pi_{1} \cap \pi_{2} }} (1-p_{e}) \), the connection's survivability level.
Here, we may employ any pseudo-polynomial time algorithm, e.g.\TR{ \cite{lawler:2001}} \cite{joksch:1966},
for solving the RSP problem.

As shall be shown, there is no solution to our problem if there is no feasible solution to the defined RSP problem. Note that each disjoint link is associated with a pair of disjoint paths in the original network, while each simple link is associated with a regular link.
Accordingly, in the third stage, we  construct the sought pair of paths of a survivable connection $(\pi_{1},\pi_{2})$ out of the links of the RSP solution, i.e. path $\tilde{\pi}$. Then, the algorithm outputs the optimal survivable solution $(\pi_{1},\pi_{2})$.

\TR{
The following theorem establishes the correctness of the CO-TSCMQ Algorithm.
\begin{theorem}
\label{theorem:algo_CO-Constrained Survivability Min-QoS Problem}
Given are a network $G(V,E)$, a pair of nodes $s$ and $t$, a survivability level constraint ${S} \in [S_{min},1]$. If there exists a survivable connection with a survivability level of at least ${S}$, then the CO-TSCMQ Algorithm returns a survivable connection that is a solution to the CO-CSMQ Problem; otherwise, the algorithm fails.
\end{theorem}

In order to prove Theorem \ref{theorem:algo_CO-Constrained Survivability Min-QoS Problem}, we begin by observing that the structure of any survivable connection $(\pi_{1},\pi_{2})$
is composed of two types of segments (see Fig. \ref{fig:segments}). The first type is a \emph{common segment}, which contains links that are common to $\pi_{1}$ and $\pi_{2}$. The second type is a \emph{disjoint segment}, which contains links that are exclusive to one of the paths. Consider a segment with a head node $u$ and a tail node $v$. A disjoint segment is a disjoint couple of paths from $u$ to $v$ denoted as $_{u}^v(\pi_{1},\pi_{2})_{ds}$.  The common segments concatenated with the disjoint segments form a survivable connection $(\pi_{1},\pi_{2})$, as illustrated in Fig. \ref{fig:segments}.
\begin{figure}[tb]
  \centering
    \includegraphics[width=0.48\textwidth]{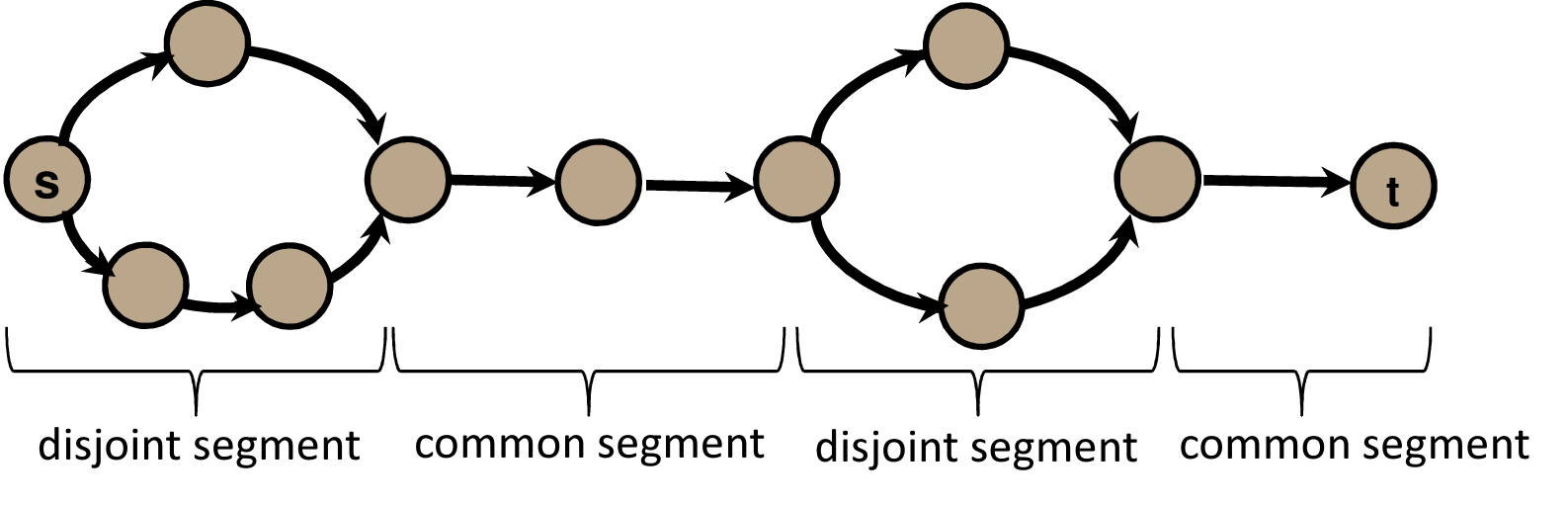}
  \caption{Concatenated common and disjoint segments forming a survivable connection}
\label{fig:segments}
\end{figure}

The \emph{weight of a disjoint segment} $W(_{u}^v(\pi_{1},\pi_{2})_{ds})$ is defined as the sum of all link weights in the disjoint segment, i.e. $W(_{u}^v(\pi_{1},\pi_{2})_{ds})= \sum_{e \in _{u}^v(\pi_{1},\pi_{2})_{ds}} w_{e} $. Accordingly, a \emph{shortest disjoint segment} is a disjoint segment $_{u}^v(\pi_{1},\pi_{2})_{ds}$ of minimum weight among all possible disjoint pairs of paths from $u$ to $v$.

\begin{lemma}
\label{lemma:disjoint_path}
Given are a network $G(V,E)$, a pair of nodes $s$ and $t$, a survivability constraint ${S} \in [S_{min},1]$. Any optimal solution $(\pi_{1},\pi_{2})$ to the respective CO-CSMQ Problem   is such that all its disjoint segments are shortest disjoint segments.
\end{lemma}

\begin{IEEEproof}
Let $(\pi_{1},\pi_{2})$ be  an optimal survivable connection of the CO-CSMQ Problem. Consider some arbitrary disjoint segment $_{u}^v(\pi_{1},\pi_{2})_{ds}$. Assume by contradiction that there is a disjoint pair of paths from $u$ to $v$, ${_{u}^v(\gamma_{1},\gamma_{2})_{ds}}$,  with a lower weight, i.e.
$W({_{u}^v(\gamma_{1},\gamma_{2})_{ds}})<W(_{u}^v(\pi_{1},\pi_{2})_{ds})$.
Due to the structure of $(\pi_{1},\pi_{2})$ survivable connections (depicted in Fig. \ref{fig:segments}), the substitution of $_{u}^v(\pi_{1},\pi_{2})_{ds}$ with
$_{u}^v(\gamma_{1},\gamma_{2})_{ds}$ forms a survivable connection from $s$ to $t$ denoted as $(\gamma_{1},\gamma_{2})$.
Note that the survivability levels of $(\gamma_{1},\gamma_{2})$ and $(\pi_{1},\pi_{2})$ are equal because
the substituted disjoint pair of paths ${_{u}^v(\gamma_{1},\gamma_{2})_{ds}}$ does not contribute to the total survivability level. Moreover,
\[W_{co,ct}(\gamma_{1},\gamma_{2})<W_{co,ct}(\pi_{1},\pi_{2})\]
which contradicts the assumption that $(\pi_{1},\pi_{2})$ is the optimal survivable connection  to the respective CO-CSMQ Problem.
\end{IEEEproof}

We recall that step 1 of the CO-TSCMQ Algorithm (Fig. \ref{algo:CO-CT}) constructs a transformed network $\tilde{G}(\tilde{V},\tilde{E})$ 
that contains ``simple links" that are the links of the original network $G(V,E)$ and ``disjoint links" that represent  disjoint shortest paths between all pairs of nodes in the network. The following two Lemmas \ref{lemma:no}
and \ref{lemma:yes} prove Theorem \ref{theorem:algo_CO-Constrained Survivability Min-QoS Problem}.
\begin{lemma}
\label{lemma:no}
If there is no solution to the RSP problem in stage 2 of the CO-TSCMQ Algorithm, then there is no solution to the CO-CSMQ Problem.
\end{lemma}

\begin{IEEEproof}
Assume by contradiction that $(\pi_{1},\pi_{2})$ is a solution to the CO-CSMQ Problem. Given a {lower} bound ${S}$, $(\pi_{1},\pi_{2})$  satisfies   $\prod_{e \in \pi_{1} \cap \pi_{2}} (1-p_{e}) \geq {S}$ and minimizes $W_{CO}(\pi_{1},\pi_{2})$. As previously mentioned, the structure of $(\pi_{1},\pi_{2})$ contains disjoint and common segments (Fig. \ref{fig:segments}). According to Lemma  \ref{lemma:disjoint_path}, the disjoint segments of $(\pi_{1},\pi_{2})$ are shortest disjoint segments. Hence, consider the path $\gamma$ that contains disjoint links equivalent to the disjoint segments and simple links equivalent to the common segments in the transformed network $\tilde{G}(\tilde{V},\tilde{E})$. 
The path $\gamma$  solves the RSP problem in stage 2, which contradicts the assumption that there is no solution to it.
\end{IEEEproof}

\begin{lemma}
\label{lemma:yes}
If there is a solution to RSP problem in stage 2 of the CO-TSCMQ Algorithm, then stage 3 of the CO-TSCMQ Algorithm  returns a solution to the CO-CSMQ Problem.
\end{lemma}
\begin{IEEEproof}
Consider the RSP solution $\gamma$ in the transformed network $\tilde{G}(\tilde{V},\tilde{E})$,
which contains disjoint links and simple links.  At stage 3, we decompose a survivable connection $(\pi_{1},\pi_{2})$  from the RSP solution $\gamma$ by transforming each disjoint link into a disjoint segment and each simple link into a link in a common segment. Given an upper bound ${S}$, $(\pi_{1},\pi_{2})$  satisfies   $\prod_{e \in \pi_{1} \cap \pi_{2}} (1-p_{e}) \geq {S}$ and minimizes {$W_{CO}(\pi_{1},\pi_{2})$}, hence it solves the CO-CSMQ Problem.
\end{IEEEproof}

\comments{
Given are a network $G(V,E)$, a pair of nodes $s$ and $t$, and a CO-weight constraint ${B} \geq 0$. We proceed to present an algorithmic scheme for solving the CO-CQMS Problem, denoted as the \emph{CO-QoS Aware Max Survivable Connection} (CO-QAMSC) Algorithm. The scheme is similar to the CO-TSCMQ Algorithm illustrated in Fig. \ref{algo:CO-CT}, except for the second stage. In this stage we swap the roles of the objective function and constraint function of the RSP problem. Namely, we solve the RSP problem with a constraint of ${B}$ while minimizing the survivability level of a path for the transformed network $\tilde{G}(\tilde{V},\tilde{E})$.
Specifically, the CO-QAMSC Algorithm finds a pair of paths that minimizes
\[ -\sum_{\substack{e \in \pi_{1} \cap \pi_{2} }} ln(1-p_{e}) = -ln \prod_{\substack{ e \in \pi_{1} \cap \pi_{2} }} (1-p_{e}) \]
and, therefore, maximizes \( \prod_{ \substack{ e \in \pi_{1} \cap \pi_{2} }} (1-p_{e}) \), the connection's survivability level.
}
}

The following theorem establishes the correctness of the CO-QAMSC Algorithm.
\begin{theorem}
\label{theorem:algo_CO-Constrained QoS Max-Survivability Problem}
Given are a network $G(V,E)$, a pair of nodes $s$ and $t$ and a co-weight constraint ${B} \geq 0$. If there exists a survivable connection with a CO-weight of at most ${B}$, then the CO-QAMSC Algorithm returns a survivable connection that is a solution to a CO-CQMS Problem; otherwise, the algorithm fails.
\end{theorem}
\TON{
\begin{IEEEproof}
See \cite{TR-QoS-Aware-Survivability}.
\end{IEEEproof}
}

\TR{
The proof is similar to that of Theorem \ref{theorem:algo_CO-Constrained Survivability Min-QoS Problem}, using the following Lemma \ref{lemma:disjoint_path2} instead of Lemma \ref{lemma:disjoint_path}.

\begin{lemma}
\label{lemma:disjoint_path2}
Given are a network $G(V,E)$, a pair of nodes $s$ and $t$, a CO-weight constraint ${B} \geq 0$.
There is an optimal solution $(\pi_{1},\pi_{2})$ to the respective CO-CQMS Problem   such that all its disjoint segments are shortest disjoint segments.
\end{lemma}

\begin{IEEEproof}
Given a  CO-CQMS Problem where ${B}$ is the QoS bound,
denote by ${S}$  the maximum survivability level of its solution. Let us define a CO-CSMQ  Problem where ${S}$ is the survivability level constraint and assume that $(\pi_{1},\pi_{2})$ is a solution to the defined problem. Clearly, the weight of the optimal solution $(\pi_{1},\pi_{2})$ is at most ${B}$, i.e. $W(\pi_{1},\pi_{2}) \le {B}$. Therefore, $(\pi_{1},\pi_{2})$ is also a solution to the CO-CSMQ Problem. According to Lemma \ref{lemma:disjoint_path}, all  $(\pi_{1},\pi_{2})$ disjoint segments are shortest disjoint segments.
\end{IEEEproof}
}

We proceed to analyze the running time of the CO-QAMSC Algorithm. As mentioned, the input size is represented by $N$ and $M$, which are the numbers of nodes and links  in the network, respectively. We denote by $R(N,M)$ and $D(N,M)$ the running time expressions of the employed (standard) RSP algorithm and EDSP algorithm, respectively.
\begin{theorem}
\label{theorem:CO-complexity}
The time complexity of the CO-QAMSC Algorithm is $O(N^2 \cdot D(N,M) + R(N, N^2))$, i.e. $O(M \cdot  N^2 + N^3 \cdot (\log(N) +  B))$.
\end{theorem}
\TON{
\begin{IEEEproof}
See \cite{TR-QoS-Aware-Survivability}.
\end{IEEEproof}
}
\TR{
\begin{IEEEproof}
Let us analyze the steps of the CO-QAMSC Algorithm, illustrated in Fig. \ref{algo:CO-CT}. At stage 1, we construct a new network from different original network links and the disjoint shortest paths between every two nodes. As each original network node and link are duplicated, the running time of first and second steps of stage 1 is $O(N)$ and $O(M)$ respectively. Next, at the third step of stage 1, we perform the Disjoint Shortest Path algorithm for each couple of nodes in total $N^2$ times and its running time is $O(N^2 \cdot D(N,M))$. At stage 2 we run the RSP algorithm in the new constructed network, which contains exactly the same number of nodes $N$ and at most $M + N^2$ links, where the complexity of this step is $R(N, N^2)$. At stage 3, we go over all the links in the new network, and its running time is $O(N^2)$. Therefore, the total complexity of the CO-QAMSC Algorithm is given by $O(M + N^2 \cdot D(N,M) + R(N, N^2) + N^2) = O(N^2 \cdot D(N,M) + R(N, N^2))$.
Now, we examine the complexity of $R(N,M)$ and $D(N,M)$.
The link-disjoint shortest pair algorithm  can be performed in $O(M + N \cdot \log(N))$ \cite{Suurballe:1984}. According to \cite{joksch:1966}, a pseudo-polynomial algorithm for the RSP problem can be performed in $O(M \cdot N \cdot B)$, where B is the constraint size.
Thus, we conclude that the CO-QAMSC algorithm is bounded by $O(M \cdot  N^2 + N^3 \cdot (\log(N) +  B) )$.
\end{IEEEproof}

The running time for the CO-TSCMQ  Algorithm can be obtained by replacing the constraint B with the desired precision of the constraint $-\ln(S)$. Note that the complexity depends linearly on the chosen precision.
}


\subsection{ Pseudo-Polynomial Schemes for Establishing CT-QoS Aware $p$-Survivable Connections}
\label{section:CT-algo}

We will now exploit the rather salient property of the optimal solutions to the CT-problems, as established in Section \ref{sec:ct_charc}. This will allow us to improve the computational complexity of the algorithmic solution of CT-CQMS and CT-CSMQ optimization problems.

{Previously, in Section \ref{sec:algo-CO-QoS}, we noted that both the CO-CQMS and the CO-CSMQ problems have quite similar solutions. Since this is the case also for the CT-problems, we shall focus on the solution to the CT-CQMS problem.}
We proceed to present an algorithmic scheme for solving the CT-CQMS Problem (Def. \ref{def:CT-CQMS}),
denoted as the \emph{CT-QoS Aware Max Survivable Connection} (CT-QAMSC) Algorithm.
{The algorithm is specified (again) in Fig. \ref{algo:CO-CT}, however now the full-line-boxed text should be disregarded while the dashed-boxed text (with gray background) should be considered.}
Moreover, in Stage 2, the text inside the double-framed-box  should be considered.

Note that the CT-Algorithmic scheme is similar to the previously presented CT-algorithmic scheme, except for two important changes{ marked by dashed-boxed text (with gray background)}. The first is in the transformation of simple links in the new constructed network in step 1.2.
Recall that simple links represent critical links of the solution, i.e., the survivable connection $(\pi_{1},\pi_{2})$.
Since in the CT problems the weight of each such link is counted twice, the weight of simple links is set to be twice the weight of the links in the original network, as illustrated {(again)} in Fig.
{\ref{fig:simple_tranformation}, where the full-line-boxed text should be disregarded now while the dashed-boxed text (with gray background) should be considered.}


The second change is the addition of a preliminary stage, namely Stage 0, to the CT algorithmic variants. At this initial stage,
the algorithm  first finds a weight-shortest path in the network $G(V,E)$ by employing a well-known shortest path algorithm, such as Dijkstra's \cite{Tarjan:1984}.
According to Theorem \ref{allin_theorem} and its corollaries, in an optimal solution of a CT problem, each of the critical links is included in any weight-shortest path.
Therefore, we can have the CT-QAMSC Algorithm focus on just nodes and links that belong to some (any) weight-shortest path. Accordingly, at Stage 1 of the algorithm, the transformed network $\tilde{G}(\tilde{V},\tilde{E})$ is limited to simple links that correspond to the weight-shortest path found at Stage 0, and to disjoint links that correspond to EDSPoPs between pairs of nodes of the identified weight-shortest path. As shall be shown, this change improves the computational complexity of the solution.

{
\comments{
Next, consider the CT-CSMQ Problem (Def. \ref{def:CT-CSMQ}), for a network $G(V,E)$, a pair of nodes $s$ and $t$, and a survivability level constraint ${S} \in [S_{min},1]$. We proceed to present an algorithmic solution scheme, termed the \emph{CT- Tunable Survivable Connection Min-QoS} (CT-TSCMQ). The scheme is similar to the CT-QAMSC Algorithm illustrated in  Fig. \ref{algo:CO-CT} except for the second stage. In that stage, we swap the roles of the objective and constraint functions of the RSP problem. Namely, we solve the RSP problem with a constraint of $-\ln({S})$ while minimizing  the weight of a path for the transformed network $\tilde{G}(\tilde{V},\tilde{E})$.
}
It is easy to verify that Theorem \ref{theorem:algo_CO-Constrained QoS Max-Survivability Problem}, established previously for the CO-TSCMQ algorithmic solution,  hold also for the CT-algorithmic solution variant. Therefore, the proof of the correctness of the CT-TSCMQ algorithm follows the same lines as for the CO-TSCMQ algorithm.
}

We denote the number of links in the identified weight-shortest path as $k$. The running time expression of the weight-shortest path algorithm is denoted as $SP(N,M)$. {As previously mentioned, $R(N,M)$ and $D(N,M)$ are the running time expressions of a standard RSP algorithm and a standard EDSP algorithm, respectively.}

\begin{theorem}
The time complexity of the CT-QAMSC Algorithm is $O(SP(N,M) + k^2 \cdot D(N,M) + R(k, k^2))$, i.e. $O( k^2 \cdot (M + N \cdot \log(N)) + k^3 \cdot B)$.
\end{theorem}
\TON{
\begin{IEEEproof}
See \cite{TR-QoS-Aware-Survivability}.
\end{IEEEproof}
}
\TR{
\begin{IEEEproof}
Let us analyze the different steps of the CT-QAMSC Algorithm specified in Fig. \ref{algo:CO-CT}.
At Stage 0, we calculate a weight-shortest path in the network, and its running time is $O(SP(N,M))$. Then, at Stage 1 it is enough to apply the Disjoint Shortest Path algorithm only between each couple of nodes in the weight-shortest path, and its running time is $O(k^2 \cdot D(N,M))$. Now, the constructed network contains $k+1$ nodes and two types of links, namely $k$ links of the weight-shortest path and at most $ (k + 1) \cdot k$ links associated with disjoint paths between each node of the weight-shortest path.
The running time of the RSP algorithm in the constructed network is $R(k,k^2)$. Hence, the total complexity of  CT-QAMSC Algorithm is given by $O(SP(N,M) + k^2 \cdot D(N,M) + R(k, k^2))$.
Now, we examine the complexity of the previously stated running time expressions, namely $SP(N,M)$, $R(N,M)$ and $D(N,M)$.
Note that the additive QoS metric is non-negative by definition. Therefore, we can use Dijkstra's algorithm in order to find SP(N,M). Dijkstra's algorithm running time is given by $O(M +N \cdot \log(N) )$ \cite{Tarjan:1984}.
As previously mentioned, the running time of the expressions $D(N,M)$ and $R(N,M)$ is bounded by $O(M +N \cdot \log(N) )$  and $O(M \cdot N \cdot B)$, respectively.
Thus, we conclude that CT-QAMSC is bounded by  $O(k^2 \cdot (M + N \cdot \log(N)) + k^3 \cdot B)$.
\end{IEEEproof}


}

According to \cite{braunstein2003optimal}, in power-law networks, which are known to be a good model for some portions of the Internet \cite{faloutsos1999power}, the number of links of
the shortest path grows proportionally to the logarithm of the number of the network nodes, i.e. $k\sim\log(N)$. Therefore, the running time of our algorithm can be significantly reduced in such networks.

\subsection{Approximation Schemes for Establishing QoS Aware Survivable Connections}

We proceed to establish Fully Polynomial Time
Approximation Schemes (FPTAS) for the considered problems.

First, we establish that an $\varepsilon$-approximation scheme for the CO-CQMS Problem  can be accomplished by employing
the previously defined CO-QAMSC Algorithm {(Fig. \ref{algo:CO-CT})} with the following change.
Consider a desired approximation ratio {$1 + \varepsilon$}, and recall the minimum survivability level  ${S}_{min}=(1-p_{min})^M$ specified
in Section \ref{sec:formulation}. In Stage 2 of the CO-QAMSC Algorithm, instead of employing a pseudo-polynomial (exact) solution scheme,
apply an (any) FPTAS of those proposed in the literature  for solving the RSP problem, e.g.~\cite{Lorenz:2001213}, with an approximation ratio of $\frac {-\ln (1 + \varepsilon)} { \ln {S}^{min}} $.
{The running time expression of the RSP Algorithm is $\hat{R}(N,M,\varepsilon)$.}
This modified algorithm shall be referred to as the F-CO-QAMSC Algorithm.

\begin{theorem}
\label{theorem:CO-QAMSC  approx}
The F-CO-QAMSC Algorithm is a FPTAS for the CO-CQMS Problem.
Specifically, the  weight of the provided connection is bounded by $B$ (as required) and its survivability level is at most $(1 + \varepsilon)$ smaller than the optimal survivability level.
The time complexity of the algorithm is bounded by $O(N^2 \cdot (M + N \cdot \log(N)) + N^3 \cdot (\log\log(N) + \frac{M}{\varepsilon }) )$.
\end{theorem}
\TON{
\begin{IEEEproof}
See \cite{TR-QoS-Aware-Survivability}.
\end{IEEEproof}
}
\TR{
\begin{IEEEproof}
Assume that ${S}^{opt}$ is the survivability level of the optimal solution of a CO-CQMS Problem, and that our algorithm finds a solution with a survivability level of  ${S}^{algo}$. In order to prove the $\varepsilon$ approximation of the F-CO-TSCMQ Algorithm, we shall show that $ \frac{{S}^{opt}}{{S}^{algo}} \leq {1 + \varepsilon}$. Since we applied a FPTAS to the corresponding RSP Problem using an approximation ratio of $\frac {-\ln (1 + \varepsilon)} { \ln {S}^{min}}$, we have that $\frac{-\ln {S}^{algo}}{-ln {S}^{opt}} \leq {1 + \frac {-\ln (1 + \varepsilon)} { \ln {S}^{min}}}$. Since ${S}^{min} \leq S^{opt} \leq 1 $, we get $\frac{-\ln {S}^{algo}}{-ln {S}^{opt}} \leq {1 + \frac {-\ln (1 + \varepsilon)} { \ln {S}^{opt}}}$. A simple manipulation in the previous formula gives $\frac{{S}^{opt}}{{S}^{algo}} \leq {1 + \varepsilon}$.
\comments{
\textbf{
Constructive}
Assume that ${S}^{opt}$ is the survivability level of the optimal solution of a CO-CQMS Problem, and that our algorithm finds a solution with a survivability level of  ${S}^{algo}$. In order to prove the $\varepsilon$ approximation of the F-CO-TSCMQ Algorithm, we shall show that $ \frac{{S}^{opt}}{{S}^{algo}} \leq {1 + \varepsilon}$.
Given is the RSP algorithm approximation ratio of $\eta$, we have that $ \frac{-\ln {S}^{algo}}{-ln {S}^{opt}} \leq {1 + \eta}$. Therefore, there is a $\eta$ which ${S}^{algo} \geq ({S}^{opt})^{1 + \eta}$. Then, according to our algorithm approximation definition, it is enough to find a $\eta$ which $ ({S}^{opt})^{1 + \eta} \geq \frac{{S}^{opt}}{1 + \varepsilon}$, i.e., $\eta \leq \frac {-\ln (1 + \varepsilon)} { \ln{{S}^{opt}}}  $. Considering $ {S}^{opt} \geq {S}^{min}$, we can define $\eta(\varepsilon) = \frac {-\ln (1 + \varepsilon)} { \ln {S}^{min}} = \frac {-\ln (1 + \varepsilon)} { M \cdot \ln (1-p_{min})}  $.
}
Given the first-order Taylor approximation we conclude that $ O(\frac {-\ln (1 + \varepsilon)} { \ln {S}^{min}}) \approx O(\frac {\ln (1 + \varepsilon)}{M}) \approx O( \frac {\varepsilon}{M})$.
The FPTAS for the RSP problem  provided by \cite{Lorenz:2001213}, assuming an approximation factor of $\eta$, has time complexity of $O(N \cdot M \cdot (\log\log(N) + \frac{1}{\eta }))$.
Thus, according to Theorem \ref{theorem:CO-complexity}, we conclude that the time complexity of the F-CO-TSCMQ  Algorithm is bounded by $O(N^2 \cdot (M + N \cdot \log(N)) + N^3 \cdot (\log\log(N) + \frac{M}{\varepsilon }) )$.
Thus, we have established a FPTAS.
\end{IEEEproof}
}

\TR{
We proceed to establish  an $\varepsilon$-approximation scheme for the CO-CSMQ Problem  by employing the previous defined  CO-TSCMQ Algorithm  with the following alteration. Given the desired approximation ratio of $\varepsilon$, in Stage 2 of the CO-TSCMQ Algorithm, apply an (any of those proposed in the literature, e.g.~\cite{Lorenz:2001213}) FPTAS for solving the RSP problem with an approximation ratio of $\varepsilon$,  instead of applying a pseudo-polynomial (exact) solution scheme.
This modified algorithm is denoted as the F-CO-TSCMQ Algorithm.
\begin{theorem}
\label{theorem:CO-TSCMQ  approx}
The F-CO-TSCMQ Algorithm is a FPTAS for the CO-CSMQ Problem. Specifically, the  survivable connection solution survivability level is bounded by $S$ and its weight is at most $(1 + \varepsilon)$ greater than the optimal weight. Moreover, the algorithm time complexity is bounded by  $O(N^2 \cdot M \cdot \log_{1+\frac{M}{N}}(N) + N^3 \cdot (\log\log(N) + \frac{1}{\varepsilon }) )$.
\end{theorem}
\begin{IEEEproof}
Assume that ${B}^{opt}$ and ${R}^{opt}$ are  the weight of the optimal solution of a CO-CSMQ Problem and the RSP problem, respectively. Moreover, assume that ${B}^{algo}$ and ${R}^{algo}$ are the weight of the solution of F-CO-TSCMQ Algorithm and the RSP algorithm of stage 2, respectively.
Given is the RSP algorithm approximation where $ \frac{R^{algo}}{R^{opt}} \leq {1 + \varepsilon}$. We note that ${B}^{opt}$ is identical to $R^{opt}$
and ${B}^{algo}$ is  equal to $R^{algo}$.
Therefore, $ \frac{{B}^{algo}}{{B}^{opt}}=\frac{R^{algo}}{R^{opt}} \leq {1 + \varepsilon}$.
Thus, according to Theorem \ref{theorem:CO-complexity}, we conclude that the time complexity of F-CO-TSCMQ  Algorithm is bounded by $O(N^2 \cdot M \cdot \log_{1+\frac{M}{N}}(N) + N^3 \cdot (\log\log(N) + \frac{1}{\varepsilon }))$. \end{IEEEproof}
}

A FPTAS for{ the CT-CSMQ Problem (Def. \ref{def:CT-CSMQ}) and} the CT-CQMS Problem{ (Def. \ref{def:CT-CQMS})} can be established by employing the same approach to the{ CT-TSCMQ and} CT-QAMSC algorithmic schemes.

\TR{
\subsection{A Numerical Example}
We further demonstrate the operation of the \TR{CO-TSCMQ and CT-QAMSC algorithms}\TON{CT-QAMSC algorithm} through an example depicted in Fig. \ref{fig:algo_execution}.

\begin{figure}

    \centering
    \begin{subfigure}[tb]{0.4\textwidth}
        \centering
        \includegraphics[width=\textwidth]{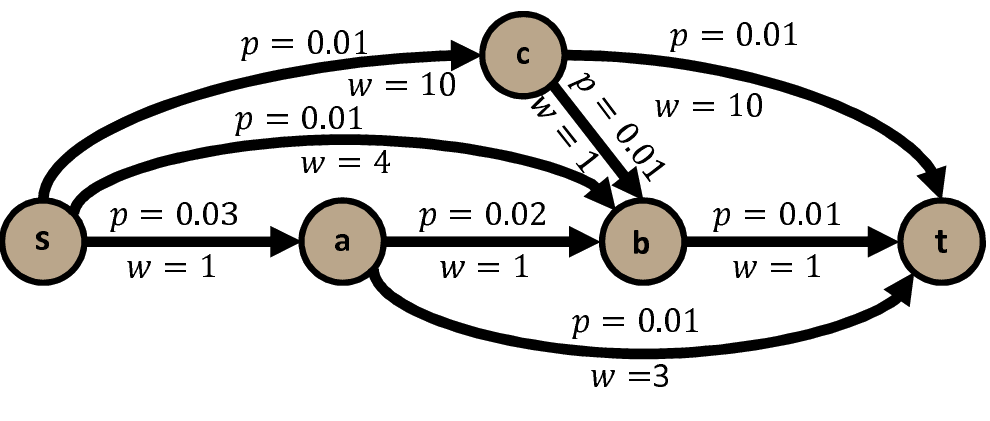}
        \caption{Network illustrating the operation of the CO-TSCMQ and CT-QAMSC algorithms}
        \label{fig:algo_example}
    \end{subfigure}%

    \begin{subfigure}[tb]{0.4\textwidth}
        \centering
        \includegraphics[width=\textwidth]{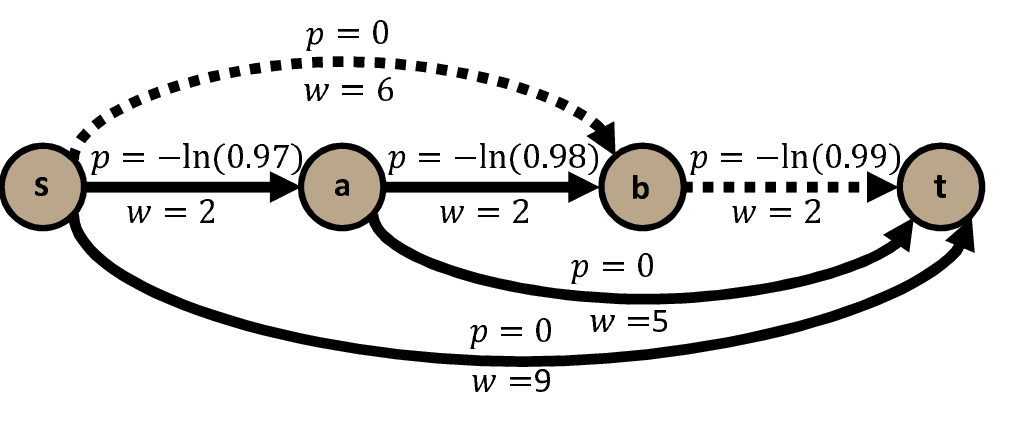}
        \caption{Transformed network of the CT-QAMSC algorithm}
        \label{fig:algo_example_ct}
    \end{subfigure}%

\TR{
    \begin{subfigure}[tb]{0.4\textwidth}
        \centering
        \includegraphics[width=\textwidth]{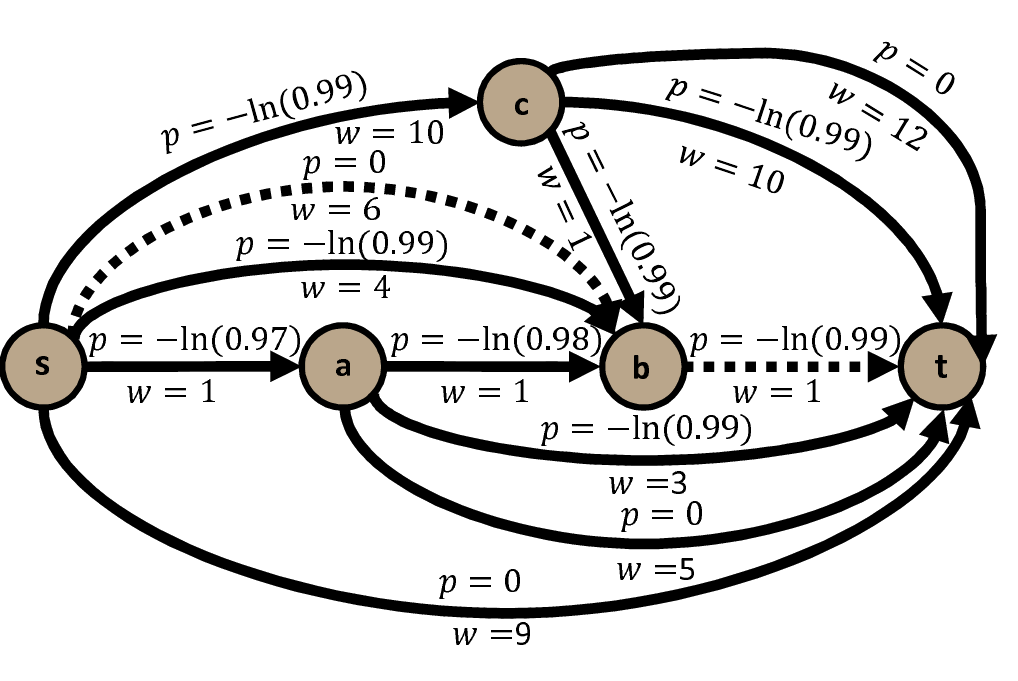}
        \caption{The transformed network of the CO-TSCMQ algorithm}
        \label{fig:algo_example_co}
    \end{subfigure}%
}

\TR{\caption{Example of the CO-TSCMQ and CT-QAMSC algorithms execution} \label{fig:algo_execution}}
\TON{\caption{Example of the CT-QAMSC algorithms execution} \label{fig:algo_execution}}
\end{figure}

Consider the network illustrated in Fig. \ref{fig:algo_example}, where the links weights $w_{e}$ and failure probabilities $p_{e}$ are depicted next to each link.
Assume that we aim at finding a  survivable connection $(\pi_{1},\pi_{2})$ with maximum survivability level and an upper-bound of $8$ on the total weight{, for the CT-CQMS problem}. As shown in Fig.{ \ref{algo:CO-CT}}\comments{ \ref{algo:CO-CT}}, the CT-QAMSC Algorithm starts by finding a weight-shortest path between the source and destination at Stage 0, which in this case is the path $ \pi_{min}=<s,a,b,t>$. Consequently, the next stage only focuses on the nodes and links of the shortest path $\pi_{min}$.
At Step $1.2$, the algorithm creates the simple links of the transformed network by duplicating the links of the shortest path $\pi_{min}$ and setting its {length} to be $l_e=-\ln(1-p_{e})$ and its time to be $t_e=2 \cdot w_{e}$.
At Step $1.3$, the algorithm finds, between each couple of nodes along the shortest path $\pi_{min}=<s,a,b,t>$, an Edge-Disjoint Shortest Pair of Paths (EDSPoP). In this specific case, three EDSPoPs are found:
the first is found between node $s$ and node $b$ $(\pi_{1}^{sb},\pi_{2}^{sb})=(<s,b>,<s,a,b>)$ with a total weight of 6, the second is found between node $a$ and node $t$ $(\pi_{1}^{at},\pi_{2}^{at})=(<a,t>,<a,b,t>)$ with a total weight of 5, and the third is found between node $s$ and node $t$ $(\pi_{1}^{st},\pi_{2}^{st})=(<s,b,t>,<s,a,t>)$ with a total weight of 9.
We create a disjoint link for each of the above three EDSPoPs, setting its lenght $l_e$ to be $0$ and its time $t_e$ to be the EDSPoP's total weight.  At the end of Stage 1, we obtain the transformed network illustrated in Fig. \ref{fig:algo_example_ct}. At stage 2, in the transformed network, the algorithm solves the RSP problem, considering a bound of $8$. Accordingly, we obtain the dashed path $\tilde{\pi}=<s,b,t>$ in Fig. \ref{fig:algo_example_ct}. Finally, at stage 3, the algorithm constructs and outputs the survivable connection $(\pi_{1},\pi_{2})=(<s,b,t>,<s,a,b,t>)$.
\TR{

Now, we consider the CO-TSCMQ Algorithm described in Fig. \ref{algo:CO-CT} given the network in Fig.  \ref{fig:algo_example}.
Assume that we aim to find a  survivable connection $(\pi_{1},\pi_{2})$ which minimizes its CO-weight and its survivability level is restricted to $0.99$. In contrast to the previous example of the CT-CQMS algorithm, Stage 1 of the CO-TSCMQ Algorithm considers all network links and nodes. At Step $1.2$, the algorithm creates the simple links of the transformed network by duplicating the original network links and setting its {length} to $l_e = w_{e}$ and its time to $t_e=- \ln(1-p_{e})$.
At Step $1.3$, the algorithm finds between each couple of nodes of the original network an EDSPoP. In this case, $4$ EDSPoPs are found: The three same EDSPoPs mentioned in the previous CT-CQMS Algorihtm example and an additional EDSPoP between node $c$ and node $t$, $(\pi_{1}^{ct},\pi_{2}^{ct})=(<c,t>,<c,b,t>)$ with a total weight of 12. Here, we consider node $c$ that does not belong to any shortest path.
As the previous example, we create a disjoint link for each found EDSPoP setting its time $t_e$ to be 0 and its {length} $l_e$ to be the EDSPoP's total weight. At the end of Stage 1, we obtain the transformed network illustrated in Fig. \ref{fig:algo_example_co}. At stage 2, in the transformed network, the algorithm solves the RSP problem with a restriction of $-\ln(0.99)$. Accordingly, we obtain the dashed path $\tilde{\pi}=<s,b,t>$ in Fig. \ref{fig:algo_example_co}. Finally, at stage 3, the algorithm constructs and outputs the survivable connection $(\pi_{1},\pi_{2})=(<s,b,t>,<s,a,b,t>)$.
}
}

\section{Simulation Study}
\label{sec:simulation}
In this section, we demonstrate the advantages of employing tunable survivability over the traditional protection (full survivability) schemes. For concreteness, we consider delay as the additive QoS metric.
Through comprehensive simulations, we compare between the minimum delay of the optimal $p$-survivable connections, where $p \in [0.9,1)$, and the minimum delay of the optimal $1$-survivable connections, the latter being obtained through pairs of edge disjoint paths. In particular, we show that, by slightly relaxing the traditional requirement of $100\%$ protection, major improvement in terms of delay is accomplished.

\begin{figure}
    \centering
    \begin{subfigure}[tb]{0.48\textwidth}
        \centering
        \includegraphics[trim = 5mm 88mm 5mm 88mm, clip,width=\textwidth]{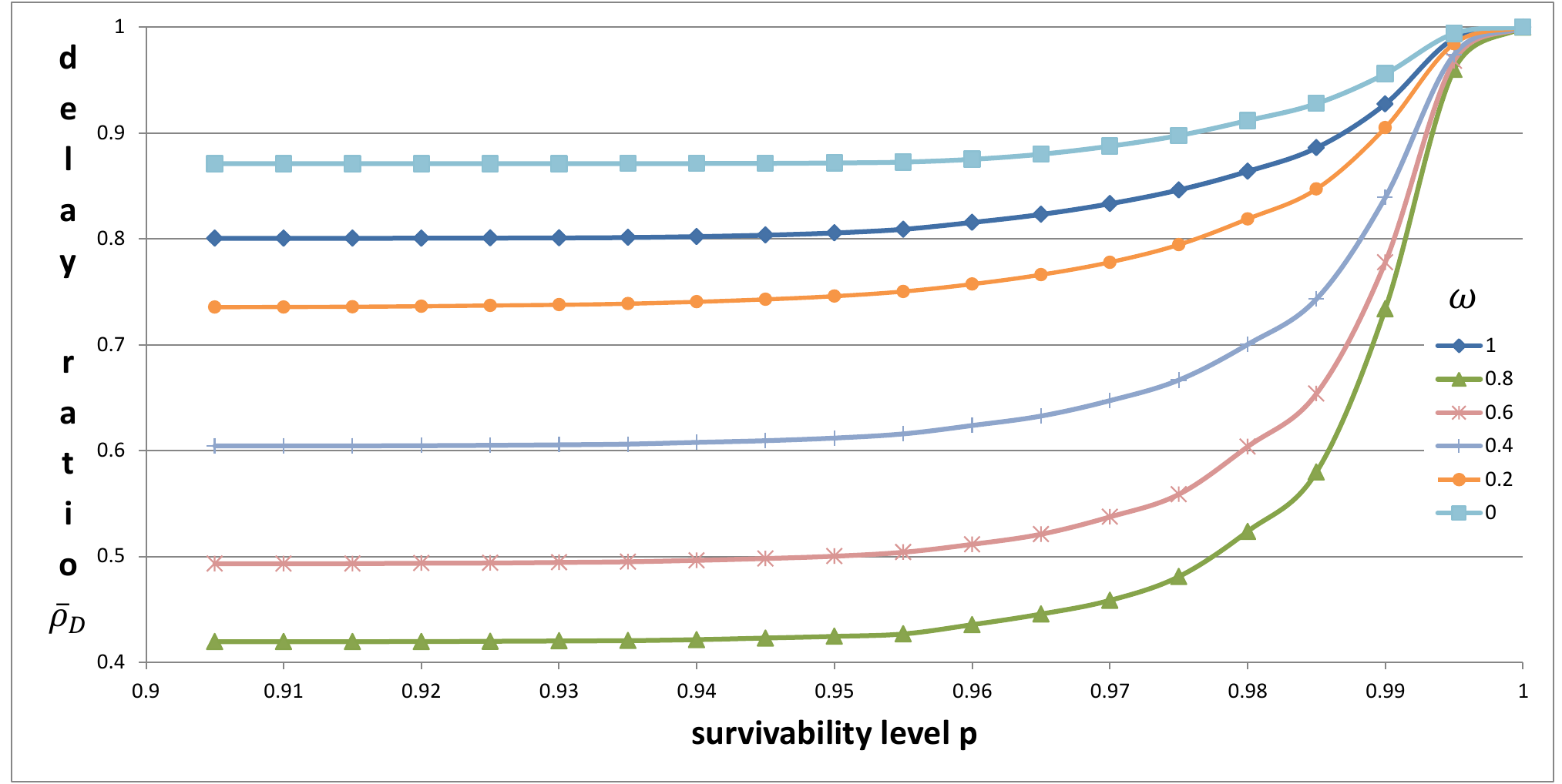}
        \caption{Power Law simulations for different values of $\omega$}
        \label{fig:PowerLaw_sim}
    \end{subfigure}%

    \centering
    \begin{subfigure}[tb]{0.48\textwidth}
        \centering
        \includegraphics[trim = 5mm 86mm 5mm 86mm, clip,width=\textwidth]{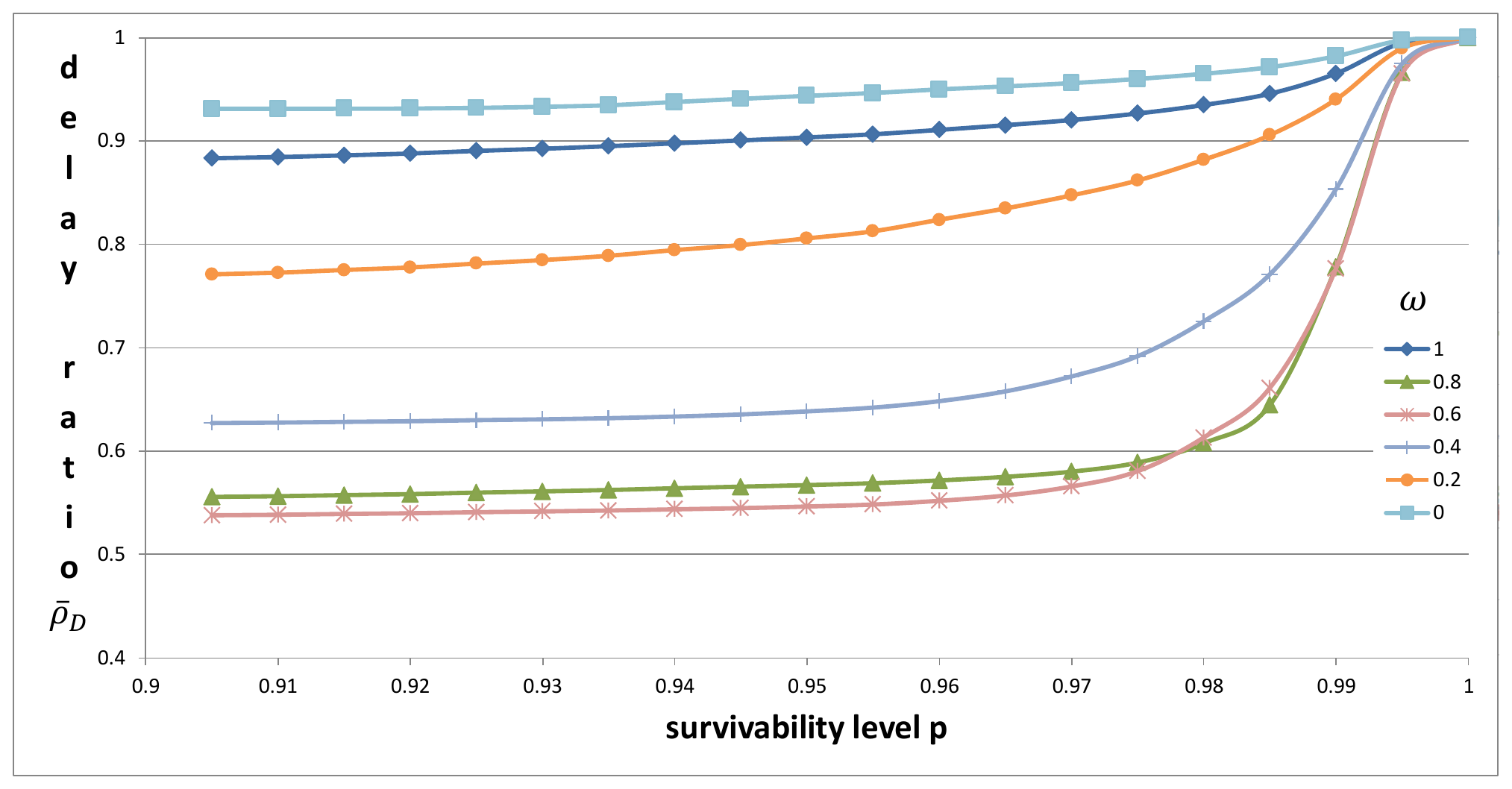}
        \caption{Waxman simulations for different values of $\omega$}
        \label{fig:Waxman_sim}
    \end{subfigure}%

\comments{
    \begin{subfigure}[tb]{0.48\textwidth}
        \centering
        \includegraphics[trim = 15mm 41mm 15mm 45mm, clip,width=\textwidth]{delay_improvement_graph.pdf}
        \caption{ Waxman vs. Power Law where  $\omega=0.7$}
        \label{fig:delay_improvement_graph}
    \end{subfigure}%
}
\caption{Average Delay Ratio versus Survivability Level}
\label{fig:sim_results}
\end{figure}

\subsection{Setup}
We generated two classes of random networks, namely Power-Law \cite{faloutsos1999power} topologies and Waxman \cite{waxman1988routing} topologies. The Power-Law topology has been shown to quite adequately model typical network interconnections, particularly, in the context of the Internet \cite{faloutsos1999power}. We demonstrate that our findings extend to other classes of network topologies by experimenting with another well known class, namely Waxman topologies.

We generated $10000$ random networks, each containing $200$ nodes, in which we identified a source-destination pair, in a manner that shall be explained later.
\comments{For ease of presentation, in this section we consider a ``reverse'' version of the CT-CQMS Problem (Def. \ref{def:CT-CQMS}), in which we minimize delay under a survivability constraint. As
mentioned, a (fully polynomial approximation) solution to this problem, termed the CT-TSMQ Algorithm, is presented in \cite{TR-QoS-Aware-Survivability}, where it is obtained  through a simple change of the above presented CT-QAMSC  Algorithm.}
In this section we consider the CT-CSMQ Problem (Def. \ref{def:CT-CSMQ}), in which we minimize delay under a survivability constraint.
For each generated network and survivability level constraint $S$ in the range of $[0.9,1]$ with intervals of $0.005$, we employed the CT-TSMQ Algorithm for the Power-Law class and for the Waxman class.
We then considered only those networks that admit $1$-survivable connections (i.e., sustain a pair of edge disjoint paths between source and destination). For each such network, we measured the minimum delay of a $p$-survivable connection, denoted as $D(p)$,  and computed the {\em delay ratio}, defined as  $\rho_{D}(p) = \frac{D(p)}{D(1)}$.
Finally, we derived the corresponding average delay ratio  $\bar{\rho}_{D}(p)$, computed over all considered network instances (of either the Power-Law or Waxman class).

In terms of delay, we considered two types of links: ``slow'' links, whose delay is set to $100$ time units,
and ``fast'' links, whose delay is set to an integer randomly (uniformly) distributed in $[1,5]$ time units. This choice represents typical mixes of links, e.g. satellite links with large  propagation delays vs. terrestrial links, or low-bandwidth (hence, large transmission delay) links vs. high-bandwidth links. Specifically, a link was classified as ''fast'' with probability of $\omega \in [0,1]$ and as ``slow'' otherwise, i.e. with probability of $1-\omega$. We ran simulations for each $\omega \in [0,1]$ value in steps of $0.2$.  The failure probability of each link was distributed normally with a mean of $1\%$ and a standard deviation of $0.3\%${, as done in \cite{Banner:2007}}.

We proceed to further specify the generation of the random topologies.
For Power-law topologies, following \cite{faloutsos1999power}, we randomly assigned a certain number of out-degree credits to each node, using the power-law distribution
\( \beta \cdot \emph{x}^{-\alpha} \), where $\emph{x}$ is a random number out of the number of network nodes,  $\alpha = 0.756 $ and  $\beta = 100$.
We connected the nodes so that every node obtained the assigned out-degree. Specifically, we randomly picked pairs of nodes $u$ and $v$, such that $u$ still had some remaining out-degree credits and then assigned a directed link $u  \rightarrow v$ between them  in case that such a link had not been assigned yet. Upon assigning such a new link, we decreased the out-degree credit of node $u$. Each simulated Power-law networks consists of $200$ nodes and in average $900$ links.

We turn to specify the generation of the Waxman topologies, following the lines of \cite{waxman1988routing}.
Initially, we located the source and the destination at the diagonally opposite corners of a square of unit dimension. Then, we randomly spread $198$ nodes over the square. Finally, for each pair of nodes $u, v$ we introduced a link $(u,v)$ with the following probability, where $\delta(u,v)$ is the distance between the nodes:
\[p(u,v) = \alpha \cdot \exp{\frac{-\delta(u,v)}{\beta \cdot \sqrt{2}}}\]
 considering $\alpha = 1.8 $ and  $\beta = 0.05$.
Each simulated Waxman network consists of $200$ nodes and in average $1800$ links.

\subsection{Results}

The simulation results are illustrated in figure \ref{fig:sim_results}. We recall that the average delay ratio
$\bar{\rho}_{D}(p)$ is a normalized metric for comparing the improvement of $p$-survivable connections over the
traditional fully disjoint path approach (i.e., $1$-survivable connections). The number of networks that admitted $1$-survivable connections was in the range of $7000$ to $8000$ (out of $10,000$), hence the samples were always significant.

The chart depicted in Fig. \ref{fig:sim_results} presents the average delay ratio improvement as a function of the
required level of survivability, for different mixes of ``fast'' and ``slow'' links (i.e., values of
$\omega$), for each of the two classes of network topologies, namely Power Law and Waxman. Overall, we observe that a modest relaxation, of a few percents in the survivability level, is enough to provide significant improvement in terms of delay.
Specifically, for Power Law networks (Fig. \ref{fig:PowerLaw_sim}),
alleviating the survivability level by about $5\%$ provides an improvement of about $20\%$ for the homogeneous case of all-fast links (i.e., $\omega=1$), and it grows to about $40\%-60\%$ for the heterogeneous cases in the range $\omega=0.4-0.8$.
Quite similar results are observed for Waxman networks (Fig. \ref{fig:Waxman_sim}).

Moreover, in all cases, most of the delay improvement is already achieved by alleviating the survivability level by about $1.5\%$.
We thus conclude that, in a typical setting, where there is some presence of relatively slower links (e.g., due to large propagation delays or low
bandwidth), a {\em modest alleviation} in the survivability level about {\em doubles the performance} in terms of delay.
\comments{
As an aside, we note that the results for $\omega$ and $1-\omega$ are not symmetric, since the delays of ``fast'' links are taken out of an interval of values whereas ``slow'' links assume a single delay value.

We have obtained similar findings for Waxman topologies. For example, for  $\omega=0.7$, a relaxation of $0.02$ in the requirement of full
survivability provides a reduction of more than $40\%$ in the end-to-end delay. The details can be found in \cite{TR-QoS-Aware-Survivability}.
}

\comments{
\begin{figure}[tb]
  \centering
    \includegraphics[width=0.48\textwidth]{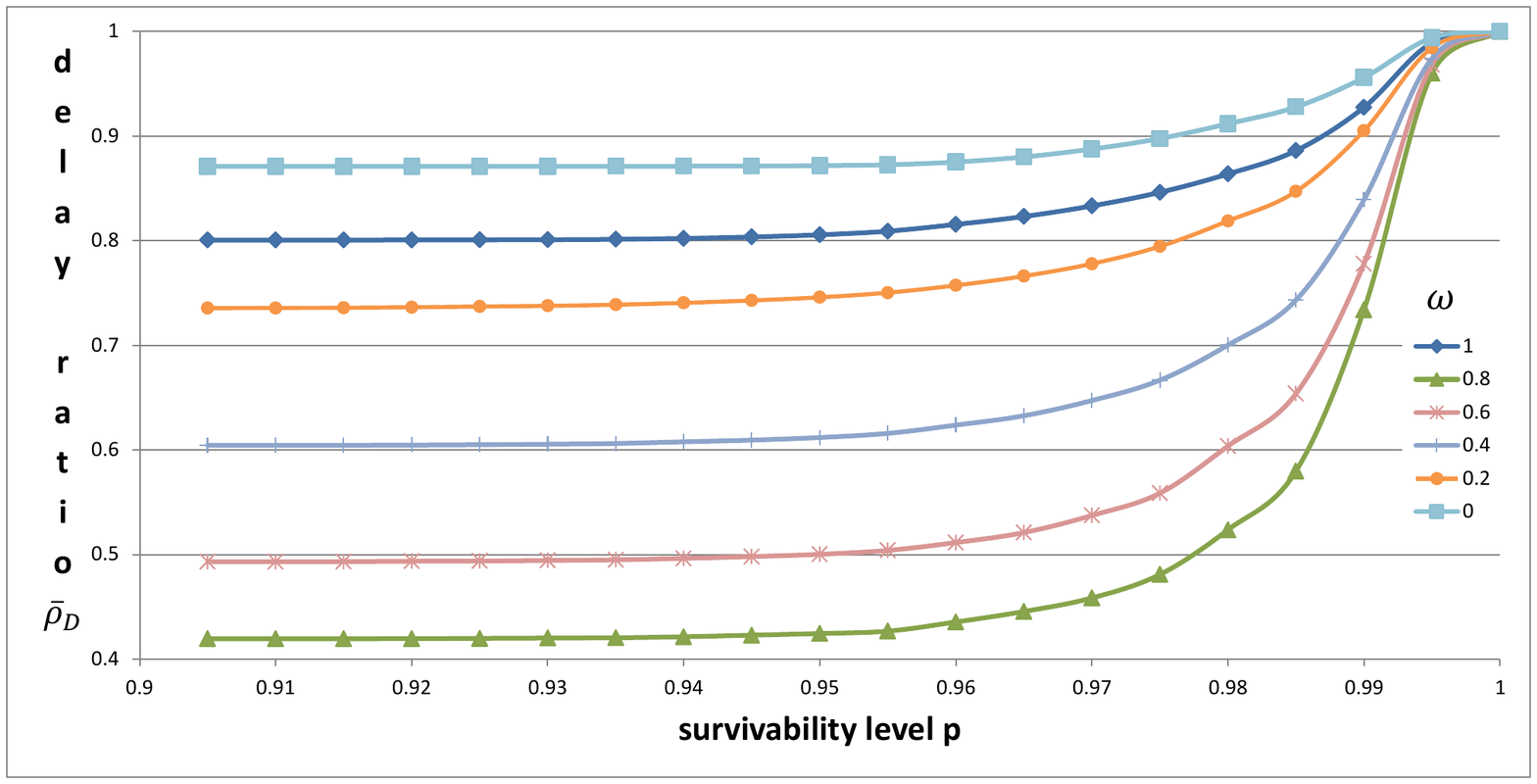}
  \caption{Delay ratio average versus the survivability level for Waxman and Power Law topologies}
\label{delay_improvement_graph}
\end{figure}
}

\section{A Network Design Perspective}
\label{sec:network_design}
Suppose that we are provided with a ``budget'' in order to improve  the total survivability level between a couple of nodes in the network, by way of upgrading the links in terms of their robustness to failures. Within this problem setting and the class of CT problems, we proceed to indicate how to exploit the particular structure of the CT solutions that has been established in section \ref{sec:ct_charc}. Theorem \ref{allin_theorem} and its corollaries significantly  reduce the amount of links that affect the optimal solution of the CT problems. Specifically, the set of candidate critical links $\mathbbm{C}$ is limited to a (typically small) subset of $E$, namely the in-all-weight-shortest-paths links $\mathbbm{L}$. This means that only these links should be considered as candidates for an upgrade.

\subsection{Discovering the in-all-weight-shortest-paths links}

We begin by sketching an algorithmic scheme for finding the in-all-weight-shortest-paths links set $\mathbbm{L}$, denoted as the \emph{In-All-Weight-Shortest-Paths Links} (IAWSPL)  \TR{Algorithm, which is illustrated  in Fig. \ref{algo_in-all-weight-shortest-paths}}\TON{Algorithm; the details can be found in \cite{TR-QoS-Aware-Survivability}}.

Given are a network $G(V,E)$ and a pair of nodes $s$ and $t$. First, our scheme finds a weight-shortest path $\pi_{short}$ between $s$ and $t$ and its weight $W(\pi_{short})$ in the original network $G(V,E)$ by employing a well-known shortest path algorithm, such as Dijkstra's \cite{Tarjan:1984}.
For each link $e$ in that weight-shortest path $\pi_{short}$, consider $\tilde{G}(\tilde{V},\tilde{E})$, which is a replica of the original network $G(V,E)$ excluding the specified link $e$.
Next, find in the network $\tilde{G}(\tilde{V},\tilde{E})$ a weight-shortest path $\tilde{\pi}_{short}$ between $s$ and $t$ and its weight $W(\tilde{\pi}_{short})$, which is, clearly, greater than or equal to $W(\pi_{short})$.
If its weight $W(\tilde{\pi}_{short})$ is greater than $W(\pi_{short})$, then the excluded link $e$ belongs to the in-all-weight-shortest-paths links set $\mathbbm{L}$.
Otherwise, i.e., if its weight  $W(\tilde{\pi}_{short})$ is equal to $W(\pi_{short})$, then the excluded link $e$ does not belong to the set $\mathbbm{L}$.
This process is repeated for all links of the weight-shortest path $\pi_{short}$ between $s$ and $t$ of the original graph $G(V,E)$.

\TR{
\begin{algorithm}[tb]
\DontPrintSemicolon
Parameters:
\Indp
$G(V,E)$- network,
$s$- source,
$t$- destination,
$\mathbbm{L}$- the in-all-weight-shortest-paths links set\;
\Indm
Variables:
\Indp
$\pi_{short}$- a weight-shortest path in the original network,
$W(\pi_{short})$ -the weight of a weight-shortest path in the original network,
$\tilde{G}(\tilde{V},\tilde{E})$- excluded link transformed network\;
$\tilde{\pi}_{short}$- a weight-shortest path in the transformed network,
$W(\tilde{\pi}_{short})$ -the weight of a weight-shortest path in the original transformed network,
\begin{enumerate}
\item Find a weight-shortest $\pi_{short}$  path between $s$ and $t$ in $G(V,E)$ and its weight is denoted by $W(\pi_{short})$.
\item For each $e \in \pi_{short}$
\begin{enumerate}
\item Set $\tilde{G}(\tilde{V},\tilde{E})$ as $G(V,E)$ excluding the link $e$
\item Find a weight-shortest path $\tilde{\pi}_{short}$ between $s$ and $t$ in $\tilde{G}(\tilde{V},\tilde{E})$  and its weight is given by $W(\tilde{\pi}_{short})$.
\item If $W(\tilde{\pi}_{short})> W(\pi_{short})$  then $\mathbbm{L}=\mathbbm{L}\cup \{e\}$. 
\end{enumerate}
\end{enumerate}
return $\mathbbm{L}$.
\caption{In-All-Weight-Shortest-Paths Links (IAWSPL) Algorithm}
\label{algo_in-all-weight-shortest-paths}
\end{algorithm}
}

\TR{
Next, we analyze the complexity of the IAWSPL Algorithm, denoting the number of links in the weight-shortest path as $K$. The IAWSPL Algorithm executes an algorithm for finding a weight-shortest path $K+1$ times. Dijkstra's algorithm can be performed for this purpose and its running time is $O(M +N \cdot \log(N))$ \cite{Tarjan:1984}. Hence, the total time complexity of the IAWSPL Algorithm is given by $O(K\cdot(M +N \cdot \log(N)))$.
}

\subsection {Optimal Links Upgrade Problem}

We proceed to formulate \TON{a network design problem that seeks}\TR{several network design problems, that seek} to allocate a given ``upgrade budget'' among the various links of the network, in a way that optimizes the total survivability level between a given pair of nodes. According to Theorem \ref{allin_theorem}, we should limit our attention only to the links that belong to the in-all-weight-shortest-paths links set $\mathbbm{L}$. Consequently, we should execute the IAWSPL Algorithm in order to find  $\mathbbm{L}$.

Given a network $G(V,E)$, each link $e \in \mathbbm{L}$  is associated with a cost  $u_e$, referred to as its \emph{upgrade level}. Accordingly, the \emph{upgrade vector} is the vector of the upgrade levels of all links in the set $\mathbbm{L}$, i.e. $U=(u_{e} | e \in \mathbbm{L} )$.\TON{
The upgrade level constitutes an additive improvement to the link's survivability level, i.e. its success probability. Such an upgrade incurs some (monetary) cost,
which is considered to be equal to the upgrade level.
Since the survivability level of any link can never exceed $1 (100\%)$, the upgrade level of a link cannot exceed $p_{e}$. We thus define the following optimization problem.
}\TR{
We consider two types of upgrade levels. In the first, the upgrade level constitutes an additive improvement to the link's survivability level, i.e.
its success probability. Such an upgrade incurs some (monetary) cost,
which is considered to be equal to the upgrade level and will never exceed $p_{e}$. In the second, the upgrade level constitutes a multiplicative improvement to the link's survivability level,
up to $\frac{p_{e} }{(1-p_{e} )}$.
We thus define the following optimization problems.
}
\begin{definition}
\label{def:Architectural_upgrading_link_Problem ver}
\textbf{\emph{Optimal Additive Upgrade Problem}}:
Given are a network $G(V,E)$, a source node $s \in V$, a destination node $t \in V$, the in-all-weight-shortest-paths links set $\mathbbm{L}$ between $s$ and $t$ and an upgrade budget $\mathbf{B}$. Each link $e\in E$ in the network is associated with a failure probability value $p_{e} \in (0,1)$. Find an upgrade vector $U=(u_{e}|e \in \mathbbm{L})$ such that:
\[
\begin{array}{l}
{\max\, \prod_{\substack{e \in \mathbbm{L} }} (1-p_{e} + u_{e} ) } \\
{s.t.}
{\sum_{\substack{e \in \mathbbm{L} }} u_{e}  \le \mathbf{B}}\\
{ \, \, \, \,\, \,\, \,\forall {e \in \mathbbm{L} }  \, \, \, \, u_{e} \ge 0}\\
{ \, \, \, \,\, \,\, \,\forall {e \in \mathbbm{L} }\, \, \, \, \, u_{e} \le p_{e}}.
\end{array}
\]
\end{definition}

\TR{
\begin{definition}
\label{def:Architectural_upgrading_link_Problem 1}
\textbf{\emph{Optimal Multiplicative Upgrade Problem}}:
Given are a network $G(V,E)$, a source node $s \in V$, a destination node $t \in V$, the in-all-weight-shortest-paths links set between $s$ and $t$  $\mathbbm{L}$ and an upgrade budget $\mathbf{B}$. Find an upgrade vector $U=(u_{e} | e \in \mathbbm{L})$ such that:

\[
\begin{array}{l}
{\max\, \prod_{\substack{e \in \mathbbm{L} }} (1 + u_{e})\cdot(1-p_{e}) } \\
{s.t.}
 {\sum_{\substack{e \in \mathbbm{L} }} u_{e}  \le \mathbf{B}}\\
{ \, \, \, \,\, \,\, \,\forall {e \in \mathbbm{L} }  \, \, \, \, u_{e} \ge 0}\\
{ \, \, \, \,\, \,\, \,\forall {e \in \mathbbm{L} }\, \, \, \, \, (1+u_{e} )(1-p_{e} )\le 1}.
\end{array}
\]
\end{definition}
}

\TR{
We proceed to establish solutions to the above problems.
\subsubsection {Optimal Additive Upgrade Problem Solution}
}
\TR{
We first note that the logarithmic operation on the objective function does not affect the additive optimization problem. Therefore, the Optimal Additive Upgrade Problem can be redefined as the following minimization problem.

\begin{equation}
\label{equation:design_problem}
\begin{array}{l}
{\min\, -\sum_{\substack{e \in \mathbbm{L} }} \ln(1-p_{e} + u_{e} ) } \\
{s.t.\,\, \, \, \, \sum_{\substack{e \in \mathbbm{L} }} u_{e} - \mathbf{B}  \le 0}\\
{\,\, \,\,\,\, \,\,\forall {e \in \mathbbm{L} }  \, \, \, \, -u_{e} \le 0}\\
{\,\, \,\,\,\, \,\,\forall {e \in \mathbbm{L} }\, \, \, \, \, u_{e} - p_{e} \le 0}.
\end{array}
\end{equation}

The above optimization problem is convex. Therefore, the \emph{Karush-Kuhn-Tucker} (KKT) conditions provide  necessary  and sufficient conditions for optimality \cite{boyd:convex}. The KKT conditions can be described as follows:

\begin{align}
  \forall{e \in \mathbbm{L}}\,\, \, \, \, -\frac{1}{(1-p_{e} + u_{e} )} +\lambda -\mu _{e} +\gamma _{e} =0\\
  {\sum_{e \in \mathbbm{L}}\,\, \, \, \,   u_{e} \le \mathbf{B}}\\
  {\forall{e \in \mathbbm{L}}\,\, \, \, \, -u_{e} \le 0}\\
  {\forall{e \in \mathbbm{L}}\,\, \, \, \,  u_{e} - p_{e} \le 0}\\
    {\forall{e \in \mathbbm{L}}\,\, \, \, \, \lambda,\mu _{e},\gamma _{e} \, \ge 0}\\
  {\lambda(\sum_{e \in \mathbbm{L}}\,\, \, \, \,   u_{e} - \mathbf{B}})=0\\
  \forall{e \in \mathbbm{L}}\,\, \, \, \,  \mu_{e} u_{e} =0\\
  {\forall{e \in \mathbbm{L}}\,\, \, \, \, \gamma_{e} ( u_{e} - p_{e} )=0}
\end{align}

The solution to this problem is

\[
u_{e}(\lambda) =\left\{
\begin{array}{l}
{p_{e}     \, \, \, \, \,\, \, \, \, \,\, \, \, \, \, \, \, \, \,\, \, \, \, \, \, \, \, \, \, \, \, \, \, \, \lambda  \le 1}\\
{\frac{1}{\lambda} - (1-p_{e})  \, \, \, \, \, \, \, \, \, 1 \le \lambda \le \frac{1}{1-p_{e}} } \\
{ 0  \, \, \, \, \,\, \, \, \, \, \, \,\, \, \, \, \, \, \, \, \, \, \, \, \, \, \, \, \, \, \, \, \, \, \, \,  \lambda > \frac{1}{1-p_{e}}  } \\
\end{array}\right.
\]

where $\lambda$ is obtained by:
\[
{\sum_{e \in \mathbbm{L}}\,\, \, \, \, \max(0,\min(\frac{1}{\lambda} - (1-p_{e}),p_e))    = \mathbf{B}}.
\]
}
\TON{
In \cite{TR-QoS-Aware-Survivability}, we show that the above optimization problem can be transformed
into an instance of the well-known Water-filling problem \cite{boyd:convex}.}
\TR{
The above optimization problem is the well-known Water-Filling problem \cite{boyd:convex}.
}
Consequently, the optimal solution is to  repeatedly split the upgrade budget among the links of the in-all-weight-shortest-paths links set
$\mathbbm{L}$  with the (currently) highest failure probability, until either the budget is exhausted or all the links assume zero failure probability.

\TON{
Furthermore, in \cite{TR-QoS-Aware-Survivability} we consider other optimization variants, e.g. a variant where the costs of upgrades are related to multiplicative (rather than additive) improvement factors.
}

\TR{
\subsubsection {Multiplicative Optimal Links Upgrade Problem solution}

We note that neither the execution of a logarithmic operation on the objective function nor the constant $(1-p_{e})$ affect the above optimization problem. Thus, the objective function can be substituted by $\sum_{\substack{e \in \mathbbm{L} }} \ln(1 + u_{e})$ and the design problem can be redefined as the following minimization problem:

\begin{equation}
\label{equation:design_problem}
\begin{array}{l}
{\min\, -\sum_{\substack{e \in \mathbbm{L} }} \ln(1 + u_{e}) } \\
{s.t.\,}
{\, \, \, \, \sum_{\substack{e \in \mathbbm{L} }} u_{e} - \mathbf{B}  \le 0}\\
{\, \, \, \,\, \, \, \,\forall {e \in \mathbbm{L} }  \, \, \, \, -u_{e} \le 0}\\
{\, \, \, \,\, \, \, \,\forall {e \in \mathbbm{L} }\, \, \, \, \, u_{e} -\frac{p_{e} }{(1-p_{e} )} \le 0}.
\end{array}
\end{equation}

In order to solve the above minimization problem, we consider the following Lagrange dual function:
\[
\begin{array}{l}
L(\bar{x},\lambda,\bar{\mu },\bar{\gamma }) = \sum _{e\in \mathbbm{L}} -\ln ( u_{e} +1)+\lambda (\sum_{e\in \mathbbm{L}}u_{e}  - B)  \\
  + \sum_{e\in \mathbbm{L}}-\mu _{e} u_{e} + \sum _{e\in  \mathbbm{L}}\gamma _{e} (u_{e}  -\frac{p_{e} }{(1-p_{e} )} ).
\end{array}
\]

In the above formulation, the objective and the inequality constraint functions are convex and affine.
Therefore, the following \emph{Karush-Kuhn-Tucker} (KKT) conditions provide  necessary  and sufficient for optimality \cite{boyd:convex}:
\begin{align}
  \forall{e \in \mathbbm{L}}\,\, \, \, \, -\frac{1}{1+u_{e} } +\lambda -\mu _{e} +\gamma _{e} =0\label{1}\\
  {\lambda(\sum_{e \in \mathbbm{L}}\,\, \, \, \,   u_{e} - \mathbf{B}})=0\\
  \forall{e \in \mathbbm{L}}\,\, \, \, \,  \mu_{e} u_{e} =0\\
  {\forall{e \in \mathbbm{L}}\,\, \, \, \, \gamma _{e} (u_{e} -\frac{p_{e} }{(1-p_{e} )} )=0}\\
  {\forall{e \in \mathbbm{L}}\,\, \, \, \, \lambda,\mu _{e},\gamma _{e} \, \ge 0}\\
  {\forall{e \in \mathbbm{L}}\,\, \, \, \, -u_{e} \le 0}\\
  {\forall{e \in \mathbbm{L}}\,\, \, \, \, u_{e} -\frac{p_{e} }{(1-p_{e} )} \le 0}\label{6}\\
  {\sum_{e \in \mathbbm{L}}\,\, \, \, \,   u_{e} \le \mathbf{B}}\label{7}
\end{align}

We proceed to establish the upgrade vector $U$ out of the above KKT conditions. From the above equations 
 we conclude that the feasible  values of $\lambda$ are in the range of $[0,1]$ and, consequently, the upgrade level of each link $u_{e}$  as a function of $\lambda$ should be:

\[
u_{e}(\lambda) =\left\{\begin{array}{l} {\frac{p_{e} }{(1-p_{e} )} \, \, \, \, \, \, \, \, \, \, \, \, \, \, \, 0 \le \lambda \le (1-p_{e} )} \\ {\frac{1}{\lambda } -1\, \, \, \, \, \, \, \, \, \, \, \, \, \, \, \, \, \, \, \, (1-p_{e} )<\lambda <1} \\ 
\end{array}\right.
\]

Consequently, there are three different cases for upgrading the desirable links. The first one is the case where the upgrade budget $\mathbf{B}$  is sufficient for fully supplying (100\%) survivability (equation \eqref{7} satisfies the strict inequality). Therefore, each link will be upgraded by  $\frac{p_{e} }{(1-p_{e} )}$. The second one is the case where
none of the links are fully upgraded (equation \eqref{6} satisfies the strict inequality for all links).  Therefore, the budget  $\mathbf{B}$ is split equally among the links in $\mathbbm{L}$. The last one is the case where the upgrade budget $\mathbf{B}$  is fully utilized but only part of the links are fully upgraded. Therefore, these  links will be upgraded by  $\frac{p_{e} }{(1-p_{e} )}$ and the remaining budget is split equally among the rest of links.

The optimal solution for the Optimal Multiplicative Upgrade Problem \ref{def:Architectural_upgrading_link_Problem 1} equally splits the budget $\mathbf{B}$
among the in-all-weight-shortest-path links of the network until a link $e \in \mathbbm{L}$ cannot be improved anymore. The algorithm,
illustrated in Fig. \ref{algo_improve}, describes the process of upgrading the various links according to the above scheme.

\begin{algorithm}[tb]

\DontPrintSemicolon
Parameters:

\Indp
$G(V,E)$- network\;
$s$- source\;
$t$- destination\;
${p_{e}}$- link failure probabilities\;
${w_{e}}$- link weights\;
$\mathbf{B}$- upgrade budget constraint\;
$U$-upgrade vector\;
\Indm
Variables:\;
\Indp
$\mathbbm{L}$- a set of links which can be improved\;
$\hat{B}$ - remaining budget\;
\Indm
Algorithm:

\begin{enumerate}
\item Initially, set $U=0$ and $\hat{B} = \mathbf{B}$.
\item Find the in-all-weight-shortest-paths links set $\mathbbm{L}$ employing IAWSPL Algorithm (Fig. \ref{algo_in-all-weight-shortest-paths}).

\While { $\mathbbm{L}$ is not empty }
{
    Find a link $e$ in $\mathbbm{L}$ with the minimum failure probability.\;
    \If { the remaining budget $\hat{B}$ can be split equally among $\mathbbm{L}$ }
    {
        \ForEach{ $e \in \mathbbm{L}$} {$u_{e} = \frac{\hat{B}}{|\mathbbm{L}|}$\;}
        \Return \;
    }
    Set  $u_{e} =\frac{p_{e} }{(1-p_{e} )}$\ and subtract the value of $u_{e}$ from $\hat{B}$\;
    Extract link e from $\mathbbm{L}$\;
}
\end{enumerate}
\caption{Optimal Links Upgrade Algorithm}
\label{algo_improve}
\end{algorithm}

Next, we analyze the time complexity of Algorithm \ref{algo_improve}, denoting the number of links in the weight-shortest path as $K$ and the number of
links in set $\mathbbm{L}$ as $L$. Algorithm \ref{algo_improve} first executes the IAWSPL Algorithm whose running time is $O(K\cdot(M +N \cdot \log(N)))$.
Next, the algorithm splits the budget among the members of the set $\mathbbm{L}$, incurring a running time of $O(L^2)$.
Hence, the total time complexity of Algorithm \ref{algo_improve} is $O(K\cdot(M +N \cdot \log(N))+ L^2)$.
}

\section{On Min-Max Survivable Connections}
\label{sec:min-max}
In this section, we proceed to consider the following variant of the survivable connections problem, termed the \emph{Constrained Survivability Min-Max-QoS (CSMMQ)} Problem, where the objective function aims at minimizing the weight of the worst of the two paths that compose  the survivable connection. We will show that any solution to the CT-CSMQ optimization problem (Def. \ref{def:CT-CSMQ}) provides a $2$-approximation scheme for the new problem variant.
We proceed to formally define the CSMMQ problem.

\begin{definition}
\label{def:CSMMQ}
\textbf{\emph{Constrained Survivability Min-Max-QoS (CSMMQ) Problem}}:
Given are a network $G(V,E)$, a source node $s \in V$, a destination node $t \in V$ and a survivability level~${S} \in [S_{min},1]$. Find a survivable connection \mbox{$(\pi_{1},\pi_{2}) \in P^{(s,t)} \times P^{(s,t)}$} from $s$ to $t$ such that:
\[
\min\,\, { W(\pi_{2})}\\
\]
\[s.t. \,\, W(\pi_{1}) \leq W(\pi_{2})\ \ \
    \,\, \prod_{e \in \pi_{1} \cap \pi_{2}} (1-p_{e}) \geq {S}.\]
\end{definition}
Note that, w.l.o.g., we may assume that $\pi_2$ is the worst path in terms of its weight, i.e.  $W(\pi_{1}) \leq W(\pi_{2})$. The CSMMQ Problem is NP-hard, since the well-known NP-hard Min-Max disjoint paths problem \cite{li1990complexity} is a special case of the CSMMQ Problem by setting $S=1$.
We proceed to show that any solution to the CT-CSMQ problem  is a $2$-approximation
for the CSMMQ Problem.

\begin{theorem}
Given are an optimal solution  $(\pi_1, \pi_2)$ for the CT-CSMQ problem such that $W(\pi_2) \geq W(\pi_1)$ and an optimal solution  $(\widetilde{\pi_1}, \widetilde{\pi_2})$  for the  CSMMQ problem such that $W(\widetilde{\pi_2}) \geq W(\widetilde{\pi_1})$. The weight of path $\pi_2$ is at most $2$-times greater than the weight of path $\widetilde{\pi_2}$, i.e.  $W(\pi_2)  \leq 2 \cdot W(\widetilde{\pi_2})$.
\end{theorem}
\begin{IEEEproof}
Since $(\pi_1, \pi_2)$ is an optimal solution for the CT-CSMQ problem, we have that $W(\pi_2) + W(\pi_1) \leq W(\widetilde{\pi_2}) + W(\widetilde{\pi_1})$. Moreover, by assumption, we have that $ W(\widetilde{\pi_2}) + W(\widetilde{\pi_1}) \leq  2 \cdot W(\widetilde{\pi_2})$ and trivially $W(\pi_2)  \leq W(\pi_2) + W(\pi_1)$. Consequently, we have that $W(\pi_2)  \leq 2 \cdot W(\widetilde{\pi_2})$.
\end{IEEEproof}

Since the CT-QAMSC Algorithm (Def. \ref{algo:CO-CT})  constitutes  a $1+\epsilon$ approximation scheme for the CT-CSMQ Problem, it is also a $2$-approximation scheme for the CSMMQ problem.

\section{Conclusions}
\label{sec:conclusions}

\emph{Tunable survivability} is a novel quantitative approach, which can be tuned to accommodate any desired level (0\%-100\%) of survivability, while alleviating the full (100\%)  protection requirement of the traditional survivability schemes. In this work, we established efficient algorithmic schemes for optimizing the level of survivability while obeying an additive end-to-end QoS constraint.
Additionally, for an important class of problems, we characterized a fundamental property, by which the links that affect the total survivability level of the optimal routing paths belong to a typically small subset. This finding gave rise to an efficient design scheme for improving
the network end-to-end survivability and, additionally, the complexity of the algorithmic scheme.
Finally, through comprehensive simulations, we demonstrated the advantage of tunable survivability over traditional survivability schemes.

\TR{
We are currently investigating the practical aspects of our findings in order to implement tunable survivability schemes in MPLS network architectures, similarly to \cite{4215794}.
Furthermore, we refer the reader to \cite{KO2008} for an extension of the tunable survivability approach that handles multiple concurrent connections.}
\TR{Moreover, similarly to \cite{johnston2011robust}, \cite{johnston2011robust} and \cite{lee2010reliability}, we consider extending our model beyond the traditional single failure and cope with multiple failures.}

\TR{
The deployment of the tunable survivability concept will be considered in the context of novel architectures such as that being designed in the FP7 ETICS (Economics and Technologies for Inter-Carrier Services) project\cite{le2010etics}.
In addition, while our work has focused on centralized algorithms, the distributed implementation of our algorithmic schemes is yet another important issue for future investigation.
}

While there is still much to be done towards the actual deployment of the tunable survivability approach,
we believe that this study provides evidence to the profitability of implementing this novel concept,
as well as useful insight and building blocks towards the construction of a comprehensive solution.

\TON{
\noindent
{\bf Acknowledgement:} This research was supported by the European Union through the ETICS project
(https://www.ict-etics.eu/) in the 7th Framework Programme, by the Israeli Ministry of Science and Technology and by the Israel Science Foundation (grant no. 1129/10).
}

\bibliographystyle{IEEEtran}
\bibliography{IEEEabrv,survivable_networks}

\TON{
\vspace{-2 cm}
\begin{IEEEbiography}[{\includegraphics[width=1in,height=1.25in,clip,keepaspectratio]{jose.jpg}}]%
{Jose Yallouz}
is a Ph.D. candidate in  the  Electrical  Engineering  department  of  the  Technion,  Haifa,  Israel.
He received the B.Sc. from  the  same  department in 2008.  He is a recipient of the Israel Ministry of Science Fellowship in Cyber and advance Computing award. He  is  mainly  interested  in  computer  networks,  algorithm  design, survivability, SDN and NFV architectures.
\end{IEEEbiography}
\vspace{-2 cm}
\begin{IEEEbiography}[{\includegraphics[width=1in,height=1.25in,clip,keepaspectratio]{Ariel_Orda_photo.jpg}}]%
{Ariel Orda}
received the B.Sc. (summa cum laude), M.Sc., and D.Sc. degrees in Electrical Engineering from the Technion in 1983, 1985, and 1991, respectively.
Since January 2014, he is the Dean of the Department of Electrical Engineering at the Technion,
where he is the Herman and Gertrude Gross Professor of Communications. His research interests include network routing, the application of game theory to computer networking, survivability, QoS provisioning, wireless networks and network pricing.
He received several awards for research, teaching and service.
\end{IEEEbiography}
}

\end{document}